\documentstyle[12pt]{article}

\begin{document}
\baselineskip=18pt plus 0.2pt minus 0.1pt
\parskip = 6pt plus 2pt minus 1pt
\newcommand{\reseteqnum}{\setcounter{equation}{0}}
\renewcommand{\theequation}{\thesection.\arabic{equation}}
\newcommand{\bx}{\mbox{\boldmath $x$}}
\newcommand{\bX}{\mbox{\boldmath $X$}}
\newcommand{\bY}{\mbox{\boldmath $Y$}}
\newcommand{\rhovev}{(\rho v)^2}
\newcommand{\mr}{\frac{g^2}{\lambda}}
\newcommand{\bra}[1]{\left\langle #1 \right|}
\newcommand{\ket}[1]{\left| #1 \right\rangle}
\newcommand{\VEV}[1]{\left\langle #1 \right\rangle}
\newcommand{\braket}[2]{\VEV{#1 | #2}}
\newcommand{\calP}{{\cal P}}
\def\Dslash{\,{\raise.15ex\hbox{/}\mkern-12mu D}}
\def\Dbarslash{\,{\raise.15ex\hbox{/}\mkern-12mu {\bar{D}}}}
\def\slashpartial{\,{\raise.15ex\hbox{/}\mkern-10.5mu \partial}}

\begin{titlepage}
\title{
\hfill
\parbox{4cm}{\normalsize KUNS-1403\\HE(TH)96/06 \\hep-th/9608060}\\
\vspace{1cm}
Multi-instanton calculus
\,\,versus\,\,
exact results \\in $N=2$ supersymmetric QCD
}
\author{Toshiyuki Harano\thanks{e-mail address: \tt
    harano@gauge.scphys.kyoto-u.ac.jp}
\, and\,\,
Masatoshi Sato\thanks{e-mail address: \tt msato@gauge.scphys.kyoto-u.ac.jp}
\\
{\normalsize\sl Department of Physics, Kyoto University, Kyoto 606-01, Japan}
\vspace{0.3cm}
}

\date{\normalsize August 1996}
\maketitle
\thispagestyle{empty}
\begin{abstract}
Microscopic tests of the exact results are performed in $N=2$
supersymmetric $SU(2)$ QCD.
We present the complete construction of the multi-instanton in $N=2$
supersymmetric QCD.
All the defining equations of the super instanton are reduced to the
algebraic equations.
Using this result, we calculate the two-instanton contribution ${\cal
  F}_2$ to the prepotential ${\cal F}$ for the arbitrary $N_f$ theories.
For $N_f=0,1,2$, instanton calculus agrees with the prediction of the
exact results, however, for $N_f=3,4$, we find discrepancies between them.
We propose improved curves of the exact results for the massive
$N_f=3$ and massless $N_f=4$ theories.
\end{abstract}

\end{titlepage}

\newpage
\section{Introduction}

In the last few years, much progress has been made in the study of the
strongly coupled supersymmetric gauge theories.
Under the holomorphy and the
duality, the low energy effective actions of $N=2$ supersymmetric Yang-Mills
theory and supersymmetric QCD in the Coulomb phase are determined exactly
for $SU(2)$ gauge group\cite{SW} and later for larger gauge
groups\cite{KLYT}-\cite{AS}.
These low energy effective theories reveal the interesting results
like the monopole condensation\cite{SW} and new supersymmetric
conformal field theories\cite{AD,APSW,EHIY} and so on.

The exact results predict the non-perturbative corrections from
instanton.
In supersymmetric theories, the dependence on the coupling constant of
the instanton corrections is completely fixed by the holomorphy and
symmetries.
Thus it is enough to calculate in the case where the coupling constant
is almost zero.
In this limit, the saddle point approximation, namely the instanton
calculus becomes exact.
Furthermore, the instanton calculus in supersymmetric
theories is safe from the infrared divergence, while the instanton
calculus in the ordinary QCD is plagued by it \cite{ADS,NSVZ,AKMRV}.
Therefore, the instanton calculus gives a reliable non-trivial test of the
exact results.
For $N=2$ supersymmetric $SU(2)$ Yang-Mills theory, various checks of the
exact result have been performed \cite{FP,DKM,Yung,FT}.
And in \cite{IS} the one-instanton calculus was extended to
$N=2$ supersymmetric $SU(N_c)$ Yang-Mills theories.
All the microscopic calculi above agree with the exact results.

The situation was changed in $N=2$ supersymmetric QCD.
The instanton calculus in $N=2$ supersymmetric $SU(2)$ QCD
was performed recently by two independent groups \cite{DKM2,AHSW,DKM3}.
A discrepancy between the instanton calculus and the exact result was
first observed in $N=2$ supersymmetric $SU(2)$ QCD with $N_f=3$
flavors \cite{AHSW}.
Since the exact result for $N_f=3$ is
derived uniquely from the exact result for the massive $N_f=4$ theory by
the decoupling argument, it also gives a discrepancy in the massive
$N_f=4$ theory.
Soon after that, for the massless $N_f=4$ theory, another discrepancies were
found in the effective coupling\cite{DKM3} and the moduli
parameter $u$\cite{AHSW}.

The aim of the present paper is to give a systematic derivation of the
results presented in our short letter\cite{AHSW} and detail
comparisons with the exact results.
In $N=2$ supersymmetric $SU(2)$ QCD, there is a parity symmetry
between hypermultiplets, and then only contributions from even number of
instanton exist \cite{SW}.
Thus the instanton corrections start from the two-instanton sector.
The construction of the supersymmetric multi-instanton in $N=2$
supersymmetric $SU(2)$ QCD was partially presented in \cite{AHSW}.
In this paper, we present the complete construction of the supersymmetric
multi-instanton in $N=2$ supersymmetric $SU(2)$ QCD and derive the
supersymmetric instanton action.
Another approach to the related topics was developed in \cite{DKM3}.
We also give some new results.
The main new results are the extension of the two-instanton calculus
to arbitrary $N_f$ theories and a proposal of improved curves for the
massive $N_f=3$ and the massless $N_f=4$ theories.

The outline of the paper is the following.
In section 2, we briefly review the exact results.
In section 3 and section 4, we construct the supersymmetric multi-instanton
in $N=2$ supersymmetric $SU(2)$ QCD.
In section 3, we derive the defining equation of the supersymmetric
instanton and the leading-order Lagrangian.
In section 4 we extend the method used in $N=2$
supersymmetric Yang-Mills theory\cite{DKM} to $N=2$ supersymmetric QCD
and find the supersymmetric instanton action.
The construction of the supersymmetric instanton is generic and not
restricted in the two-instanton sector.
In section 5, we perform the two-instanton calculus in $N=2$
supersymmetric $SU(2)$ QCD with arbitrary flavors.
In section 6, we compare the instanton calculus and the exact result.
It is found that there exist discrepancies between them.
In section 7, we summarize our conclusion and discuss the improvements
of the curves of the exact result.
In appendix A, B and C we give our conventions, the Lagrangian of
$N=2$ supersymmetric QCD and the supersymmetric transformation respectively.
In appendix D, the regularization scheme of the instanton calculus is
discussed.

\section{Brief review of the exact results}
In this section, we briefly review the exact results derived by
Seiberg and Witten \cite{SW}.
We consider the Coulomb phase of $N=2$ supersymmetric $SU(2)$ QCD.
In the generic point of the Coulomb branch,
the gauge symmetry breaks spontaneously to $U(1)$ and all
hypermultiplets are massive, and therefore the light degree of freedom
which survives
in the low energy effective theory is the $U(1)$ vector multiplet.
The holomorphy determines the low energy effective
Lagrangian for the massless vector multiplet
except for the prepotential ${\cal F}(A)$;
\begin{eqnarray}
  \label{efflag}
{\cal L}_{eff}=\frac{1}{4\pi}{\rm Im}\left[\int d^4\theta \frac{\partial
      {\cal F}(A)}{\partial A}\bar{A}+\int d^2 \theta
    \frac{\partial^2{\cal F}(A)}{\partial
      A^2}W^{\alpha}W_{\alpha}\right],
\end{eqnarray}
where $A$ and $W_{\alpha}$ are the $N=1$ chiral superfields
in the $N=2$ $U(1)$
vector multiplet.
In the semiclassical region, the prepotential are
expanded by,
\begin{eqnarray}
   {\cal F}(a)=\frac{i
    a^2}{4\pi}\left\{(4-N_f)\ln
\left(\frac{a^2}{\tilde{\Lambda}_{N_f}^2}\right)
+\sum_{k=0}^{\infty}{\cal F}_k(N_f)
\left(\frac{\tilde{\Lambda}_{N_f}}{a}\right)^{(4-N_f)k}
\right\},
\end{eqnarray}
where we define $a=\VEV{A}/2$.
The first term is the one-loop correction and the others are
instanton corrections. The coefficients
${\cal F}_{2n+1}$ vanish for $N_f > 1$ by the parity symmetry.

Seiberg and Witten derive the exact form of the prepotential, based on
the physical conjecture: the duality and the physical interpretation to
singularities of the moduli space, which is parameterized by a gauge
invariant parameter $u=\VEV{{\rm tr}\phi^2}$. They introduce
$a_D=\partial {\cal F}/\partial a$, which is related by $N=2$
supersymmetry to the dual photon. The pair ($a_D,a$) is a holomorphic
section of an $SL(2,Z)$ bundle over the punctured complex $u$-plane. The
exact results are given as the period on the torus of the holomorphic
differential. For the massless theories, the tori are given by the
elliptic curves,
\begin{eqnarray}
  \label{ellipticSYM}
 N_f=0&:& y^2=x^2(x-u)+\frac{1}{4}\tilde{\Lambda}_0^4 x,\\
  \label{ellipticSQCD}
 N_f=1,2,3&:&
y^2=x^2(x-u)-\frac{1}{64}\tilde{\Lambda}_{N_f}^{2(4-N_f)}(x-u)^{N_f-1},
\end{eqnarray}
and their results are given by,
\begin{eqnarray}
\frac{da(u)}{du} = \frac{\sqrt{2}}{8\pi}  \oint_{\alpha} \frac{dx}{y} ,
\end{eqnarray}
\begin{eqnarray}
\frac{da_D(u)}{du} = \frac{\sqrt{2}}{8\pi}  \oint_{\beta} \frac{dx}{y} .
\end{eqnarray}
By calculating the inverse function of $a(u)$, we obtain the moduli
$u(a)$. Inserting $u(a)$ to $a_D(u)$, we get the exact form
of the prepotential. The exact results predict all the series of
multi-instanton corrections to the low energy effective Lagrangian and
the moduli $u$. We obtain lower order coefficients,
\begin{eqnarray}
\label{coeff}
  \begin{array}{ccc}
   N_f & {\cal F}_1& {\cal F}_2\\
    0  &   -2^{-4} &    -5\cdot 2^{-13}\\
    1  &    0      &     3\cdot 2^{-12}\\
    2  &    0      &    -2^{-11}\\
    3  &    0      &    -2^{-10}.
  \end{array}
\end{eqnarray}
The coefficients of the higher order corrections are obtained systematically
by Picard-Fuchs equation \cite{IY}.
There is the following relation between the moduli and prepotential
\cite{Matone,STY,EY},
\begin{eqnarray}
  u(a)&=&\frac{8\pi i}{4-N_f}\left({\cal F}(a)-\frac{1}{2}a
    \partial_a{\cal F}(a)\right)
\nonumber
\\
\label{u(a)}
&=&2 a^2\left\{1-\frac{1}{2}\sum_{k=1}^{\infty}k{\cal
    F}_k(N_f)\left(\frac{\tilde{\Lambda}_{N_f}}{a}\right)^{(4-N_f)k}\right\}.
\end{eqnarray}
In the case when $N_f$=4, Seiberg and Witten assert that the quantum
theory has exact scale invariance, and that
both the moduli $u$ and prepotential  ${\cal F}$ receive no quantum
correction,
\begin{eqnarray}
u=2a^2,\quad {\cal F}=\frac{1}{2}\tau a^2,
\end{eqnarray}
where $\tau=\theta/\pi +8\pi i/g^2$ is the classical coupling constant.

When the hypermultiplets have non-vanishing bare masses, the torus is
deformed by the bare masses.
According to the decoupling relation: $m_{N_f}^2
\tilde{\Lambda}^{8-2N_f}_{N_f} =\tilde{\Lambda}^{8-2(N_f-1)}_{N_f-1}$,
the elliptic curves corresponding to different flavors are related
each other. For example, the family of the elliptic curves for the
massive $N_f=3$ theory is given by,
\begin{eqnarray}
  \label{ellipticmassive}
&&y^2=x^2(x-u)
-\frac{1}{64}\tilde{\Lambda}_3^2(x-u)^2
-\frac{1}{64}(m_1^2+m_2^2+m_3^2)
\tilde{\Lambda}_3^2 (x-u)
\nonumber
\\
&&\hspace{10ex}+\frac{1}{4}m_1m_2m_3\tilde{\Lambda}_3
x-\frac{1}{64}(m_1^2m_2^2+m_2^2m_3^2 +m_1^2m_3^2)\tilde{\Lambda}_3^2\,  .
\end{eqnarray}
By considering the decoupling limit and the massless limit,
Eq.(\ref{ellipticSYM}) and (\ref{ellipticSQCD}) are easily derived from
Eq.(\ref{ellipticmassive}).
Since there is no discrete symmetry in the moduli space for
massless $N_f=3$,  $u$ is not completely determined  by studying
only the massless $N_f=3$ theory. The freedom of adding a constant to
$u$ remains.
  We will find that this constant is determined to be zero by
the decoupling argument of the massive $N_f=4$ theory.
The massive $N_f=4$ curve is given by,
\begin{eqnarray}
&&    y^2=(x^2-c_2^2u^2)(x-c_1 u)-c_2^2(x-c_1u)^2\sum_i
  m_i^2-c_2^2(c_1^2-c_2^2) (x-c_1
  u)\sum_{i>j}m_i^2m_j^2
\nonumber
\\
\label{elliptic4}
&&\hspace{8ex}+2c_2(c_1^2-c_2^2)(c_1x -c_2^2u)m_1m_2m_3m_4
  -c_2^2(c_1^2-c_2^2)^2\sum_{i>j>k}m_i^2 m_j^2 m_k^2,
\end{eqnarray}
where $c_1=\frac{3}{2}e_1$ and $c_2=\frac{1}{2}(e_3-e_2)$ and
$e_i$
are the roots of the cubic polynomial: $4 x^3 -g_2(\tau)x-g_3(\tau)$.
Here $g_2$ and $g_3$ are defined by
$g_2=60\pi^{-4}G_4(\tau)$, $g_3=140\pi^{-6}G_6(\tau)$ and  $G_4$, $G_6$
are the Eisenstein series:
\begin{eqnarray}
  G_4(\tau)=\sum_{m,n \in Z_{\neq 0}} \frac{1}{(m\tau +n)^4},\quad
 G_6(\tau)=\sum_{m,n  \in Z_{\neq 0}} \frac{1}{(m\tau +n)^6}.
\end{eqnarray}
The roots $e_i$ obey the following equations,
\begin{eqnarray}
&&e_1+e_2+e_3=0,\\
&&e_1-e_2=\theta_3^4(0,\tau),\\
&&e_3-e_2=\theta_1^4(0,\tau),\\
&&e_1-e_3=\theta_2^4(0,\tau),
\end{eqnarray}
where $\theta_i$ are the $\theta$ functions,
\begin{eqnarray}
&&\theta_1(0,\tau)=\sum_{n\in Z} q^{(n+1/2)^2/2},\\
&&\theta_2(0,\tau)=\sum_{n\in Z} (-1)^n q^{n^2/2},\\
&&\theta_3(0,\tau)=\sum_{n\in Z} q^{n^2/2}.
\end{eqnarray}
We define $q$ by $q=\exp(2\pi i \tau)$.
In the classical limit: $\tau \rightarrow i\infty$,
\begin{eqnarray}
&&e_1=\frac{2}{3}+16 q+O(q^2),\\
&&e_2=-\frac{1}{3}-8q^{1/2}+O(q),\\
&&e_3=-\frac{1}{3}+8q^{1/2}+O(q).
\end{eqnarray}
By taking the decoupling limit: $\tau \rightarrow
i\infty$, $m_4\rightarrow \infty$ with $m_1$, $m_2$, $m_3$ and
$\Lambda_3 =64 q^{1/2} m_4$ fixed, the massive $N_f=3$ curve
(\ref{ellipticmassive}) is derived uniquely and therefore the moduli
$u$ for $N_f=3$ is determined without
 the freedom of adding a constant.

\reseteqnum
\section{Defining equations of the supersymmetric instanton}
\label{sec-def}

In the following two sections, we will construct the supersymmetric
multi-instanton in the Coulomb phase of $N=2$ supersymmetric QCD.
In the Coulomb phase,
the adjoint scalar and the gauge boson become massive, therefore
the scale invariance is broken.
It is well-known that when the scale invariance is broken,
instanton ceases to exist as a solution of the
equation of motion except for the zero-radius one.
In the formal manner, we must extend the equation of motion in order
to incorporate the instanton effects \cite{Affleck, AHSW2}.
However, when the coupling constant is weak enough, the structure of
the dominating configuration of the path integral does not depend on
the details of the extension of the equation.
In the weak coupling theory, the small size configuration dominates the
path integral, since the size of the dominating configuration $\rho$
is given by $\rho\sim g/M$, where $g$ is the coupling constant and $M$
is the Higgs mass.
Because the scale invariance is effectively
restored for the small size configuration, the dominating
configuration satisfies the classical equation in the leading order of
$g$ except for the large-range behavior, which is not relevant in the
following calculation.
In supersymmetric theories, the coupling dependence of
the instanton contribution is completely fixed, then it is sufficient
to consider the case in which the coupling constant is almost zero.
Therefore, we will solve the equation of motion in the
leading order.

In non-supersymmetric theory, the leading-order equation was already
given by 't Hooft \cite{tHooft}.
To extend to the supersymmetric theory, we must take into account
the source term given by the fermionic zero modes.
For example, the leading equation of the $N=2$ vector multiplet becomes
\begin{eqnarray}
\label{eqofmotionA}
&&\hspace{4.5ex}F_{\mu\nu}=-\tilde{F}_{\mu\nu},\nonumber\\
&&\hspace{-0.5ex}\Dbarslash\lambda=0,\hspace{6ex}\Dbarslash\psi=0,\\
&&D^2\phi-\sqrt{2}ig[\lambda,\psi]=0\nonumber.
\end{eqnarray}
Comparing to the non-supersymmetric theory, a difference appears in
the equation of the scalar field, which is given by $D^2\phi=0$ in
non-supersymmetric theory.
To clarify this, we estimate the  coupling constant dependence of the
source term.
The coupling constant dependence of instanton
solution is given by  $A_{\mu} \sim O(g^{-1})$
and because of  $N=1$ supersymmetry, we set the normalization of the
fermionic zero mode so that $\lambda \sim
O(g^{-1})$
\footnote{$\bar{\lambda}$ is not
  generated by the super transformation of $A_{\mu}$;
  $\delta{\bar\lambda}\propto\sigma_{\mu\nu}F_{\mu\nu}=0$
}.
Again because of $N=1$ supersymmetry, $\phi$ and $\psi$ have the same
coupling constant dependence, therefore, the source term of the
equation of ${\phi}$ has the same order as $D^2\phi$.
This is the reason why the 't Hooft equation must be modified in the
supersymmetric theory.
The coupling constant dependence of the $\phi$ and $\psi$ is determined
by the boundary condition of the $\phi$.
In the Coulomb branch, $\phi$ must satisfy the boundary condition
$\phi\rightarrow\VEV{\phi}$ at $x\rightarrow\infty$.
The vacuum expectation value $\VEV{\phi}$ does not depend on the coupling
constant, then $\phi,\,\psi\sim O(1)$
\footnote{We have
respected only $N=1$ supersymmetry for  simplicity. Even if we
take into account $N=2$ supersymmetry, the conclusion is not changed.}.

In the similar way, we can derive all the leading-order
equation.
In $N=2$ supersymmetric QCD, there appear $N_f$ hypermultiplets.
The $N=1$ fundamental chiral multiplet in the $N=2$ hypermultiplet is
characterized by the following equations:
\begin{eqnarray}
  \label{eqofmotionq}
&&\Dbarslash q=0,\hspace{15ex}\Dbarslash \tilde{q}=0,\\
  \label{eqofmotionQ}
&&D^2Q-\sqrt{2}ig\lambda q=0,\hspace{3ex}
D^2\tilde{Q}+\sqrt{2}ig\tilde{q}\lambda=0.
\end{eqnarray}
As well as the vector multiplet, because of the $N=1$ supersymmetry,
a source term appears in (\ref{eqofmotionQ}).
As will be shown below, if we demand that the kinetic term
of $q$ and $\tilde{q}$ is the same order as that of $\lambda$, the
normalization of $q$ and $\tilde{q}$ is determined uniquely:
$q,\,\tilde{q}\sim O(g^{-1/2})$ and $Q,\,\tilde{Q}\sim O(g^{-1/2})$.
By the $SU(2)_R$ symmetry
\begin{eqnarray}
\lambda\rightarrow\psi,
\quad Q\rightarrow\tilde{Q}^{\dagger},
\quad\tilde{Q}\rightarrow -Q^{\dagger},
\end{eqnarray}
the following equation of the fundamental
anti-chiral multiplet is derived,
\begin{eqnarray}
D^2Q^{\dagger}-\sqrt{2}ig\tilde{q}\psi=0,\quad
D^2\tilde{Q}^{\dagger}-\sqrt{2}ig\psi q=0.
\label{eqofmotiondagQ}
\end{eqnarray}
{}From this equation, we find that $Q^{\dagger},\,
\tilde{Q}^{\dagger},\,\bar{q},\,\bar{\tilde{q}} \sim O(g^{1/2})$.

In $N=2$ supersymmetric QCD, the leading-order equation of the
adjoint anti-scalar field is also modified,
\begin{eqnarray}
D^2\phi^{\dagger a}-\sqrt{2}ig\tilde{q}T^a q=0.
\label{def-dagphi}
\end{eqnarray}
The source term comes from the superpotential
$-i\sqrt{2}g\tilde{Q}\Phi Q$.
As well as $\phi$,
the boundary condition determines
the order of $\phi^{\dagger}$: $\phi^{\dagger}\sim O(1)$.
Thus the order of super partner
$\bar{\psi}$ becomes $\bar{\psi}\sim O(1)$.
Note that the source term in (\ref{def-dagphi}) is the same order as
$D^2\phi^{\dagger a}$.

Since the equations of the remaining fields are not needed in the
following calculation, we do not write them down explicitly.
By examining the classical equation of motion carefully, we obtain the
order of the remaining fields;

\begin{eqnarray}
\begin{array}{cc}
\bar{\lambda}\sim O(g),& D\sim O(g),\\
F_Q, F_{\tilde{Q}} \sim O(g^{3/2}),& F^{\dagger}_Q,
F^{\dagger}_{\tilde{Q}}\sim O(g^{1/2}),\\
F_{\phi}\sim O(g^2),&F^{\dagger}_{\phi}\sim O(1).
\end{array}
\end{eqnarray}

{}From the order of the fields obtained above, we find the
leading-order parts of the Lagrangian ${\cal L}_0$,
\begin{eqnarray}
&&{\cal L}_0=
\frac{1}{2}{\rm tr}(F_{\mu\nu}F_{\mu\nu})\nonumber\\
&&\hspace{3ex}+{\rm tr}\left\{-2i\bar{\lambda}\Dbarslash\lambda
-2i\bar{\psi}\Dbarslash\psi+2(D_{\mu}\phi)^{\dagger}D_{\mu}\phi
+2\sqrt{2}ig\lambda[\psi,\phi^{\dagger}]\right\}
\nonumber\\
&&\hspace{3ex}+(D_{\mu}Q)^{\dagger}D_{\mu}Q
+D_{\mu}\tilde{Q}(D_{\mu}\tilde{Q})^{\dagger}
 -i\bar{q}\Dbarslash q -i\bar{\tilde{q}}\Dbarslash\tilde{q}\\
\nonumber
&&\hspace{3ex}+\sqrt{2}ig\left(\tilde{q}\phi q+Q^{\dagger}\lambda
  q-\tilde{q}\lambda\tilde{Q}^{\dagger}
+\tilde{q}\psi Q+\tilde{Q}\psi q\right).
\end{eqnarray}
We have neglected $O(g^2)$ terms in the original Lagrangian.
The first term of the leading-order Lagrangian is $O(g^{-2})$ and the remaining
terms are $O(1)$.
It can be easily seen that the kinetic terms of $q$ and $\tilde{q}$ is
the same order as that of $\lambda$.

As a consistency test, we examine the Euler-Lagrangian equations of
the leading-order Lagrangian ${\cal L}_0$.
Except for the gauge field, all the leading-order equations are derived
from ${\cal L}_0$.
For the gauge field, the leading-order equation
$D_{\mu}F_{\mu\nu}=0$ is not derived from
${\cal L}_0$, but there appear source terms.
However, since it can be shown easily that these source terms
give only
$O(g^2)$ corrections in the instanton action, we can neglect these
terms in the defining equation of $F_{\mu\nu}$.
We dub the dominating configuration defined by
Eq.(\ref{eqofmotionA}), (\ref{eqofmotionq}), (\ref{eqofmotionQ}),
(\ref{eqofmotiondagQ}) and (\ref{def-dagphi}) as supersymmetric
instanton in $N=2$ supersymmetric QCD.

\reseteqnum
\section{Solutions of the defining equations}

In this section, we solve the defining equations and give the
supersymmetric instanton explicitly.
We extend the result in $N=2$ supersymmetric Yang-Mills
theory\cite{DKM} to $N=2$ supersymmetric QCD.
As we will show below, all the defining equations reduce to
the algebraic equations.
In the following, we omit the coupling constant $g$ for simplicity.
\subsection{vector multiplet}
\subsubsection{spin 1}
In this paper, we consider the gauge field $A_{\mu}$ satisfying the
anti-self-dual condition:
\begin{eqnarray}
F_{\mu\nu}=-\tilde{F}_{\mu\nu}.
\end{eqnarray}
Anti-self-dual solutions of arbitrary number of instanton are
constructed by the ADHM construction \cite{ADHM,CWS}.
We will treat the $k$-instanton in this section.
To construct the solution, we introduce a $(k+1)\times k$ matrix $M(x)$
made up of quaternions\footnote{See appendix A. The matrix representation is
given there.}.
The matrix $M(x)$ is chosen to be linear in $x$,
\begin{eqnarray}
M(x)=B-Cx,
\end{eqnarray}
where $B$ is constant $(k+1) \times k$ quaternionic matrices of
rank $k$ and $x$ is the quaternion.
We denote the elements of $B$ by a $k$-dimensional row vector $\omega$ and
a $k \times k$ matrix $\hat{a}$,
\begin{eqnarray}
B=\left(
\begin{array}{ccc}
\omega_1&\cdots&\omega_k\\
&\hat{a}&
\end{array}
\right).
\label{B}
\end{eqnarray}
The $(k+1)\times k$ matrix $C$ is chosen to the following canonical form,
\begin{eqnarray}
C=\left(
\begin{array}{ccc}
0&\cdots&0\\
&\hat{C}&
\end{array}
\right),
\quad\hat{C}_{i,j}=\delta_{i,j}
,\quad i,j=1,\cdots,k.
\end{eqnarray}
Furthermore,
\begin{eqnarray}
R(x)=M^{\dagger}(x)M(x)
\end{eqnarray}
is assumed to be a real, invertible $k\times k$ matrix:
$\bar{R}_{i,j}=R_{i,j}$.
{}From this we obtain $C^TM_{\mu}=M_{\mu}^TC$ which is useful in the
calculation.
As we will see soon, the reality condition of $R$ is equivalent to the
anti-self-duality of $F_{\mu\nu}$.

The gauge field $A_{\mu}$ is given by a quaternionic $(k+1)$-dimensional column
vector $N(x)$:
\begin{eqnarray}
{A_{\mu}}^{\dot{r}}_{\,\,\dot{s}}(x)
=iN^{\dagger\dot{r}r}(x)\partial_{\mu}N(x)_{r\dot{s}}.
\label{gauge filed}
\end{eqnarray}
$N(x)$ is given by the solution of the following algebraic equation:
\begin{eqnarray}
N^{\dagger}(x)M(x)=0,\quad N^{\dagger}(x)N(x)=1  .
\end{eqnarray}
The first equation gives $k$ quaternionic conditions on the $k+1$ elements
in $N(x)$.
The second equation determines the normalization of $N(x)$.
Thus there remains an ambiguity of $N(x)$; $N(x)\rightarrow N(x)u(x)
$, where $u(x)$ is a quaternion of unit length, $u^{\dagger}(x)u(x)=1$.
In the matrix representation of quaternion, $u(x)$ is a unitary matrix.
This is a gauge symmetry.
Applying an appropriate gauge transformation, we set the boundary
condition of the 0-th component of $N(x)$ as $N_{0}(x)\rightarrow 1$
at $x\rightarrow\infty$.
This is the multi-instanton version of the singular gauge condition.
In the singular gauge, $N$ is given by,
\begin{eqnarray}
N=\left(
\begin{array}{c}
N_0\\
N_{0+i}
\end{array}
\right)=\left(
\begin{array}{c}
\sqrt{1-\omega_iR^{-1}_{i,j}\bar{\omega}_j}\\
-M_{0+i,j}R^{-1}_{j,h}\bar{\omega}_hN_0/|N_0|^2
\end{array}
\right),
\end{eqnarray}
where $i=1,\cdots,k$.
In the singular gauge, the 0-th component plays a special role.
To distinguish the 0-th component from others clearly, we use
the index $0+i$ to denote the other components.
It is easily seen that $A_{\mu}\rightarrow O(1/x^{3})$ at
$x\rightarrow\infty$.

The following formula is useful in the multi-instanton calculus:
\begin{eqnarray}
I-N(x)N^{\dagger}(x)=M(x)R^{-1}(x)M^{\dagger}(x),
\label{uf}
\end{eqnarray}
where $R^{-1}$ is the $k\times k$ inverse matrix to $R$ and $I$ is the
$k\times k$ unit matrix.
This formula is derived from an identity
$(I-NN^{\dagger}-MR^{-1}M^{\dagger})M=0$.
{}From this identity, we find that
$I-NN^{\dagger}-MR^{-1}M^{\dagger}\propto N^{\dagger}$.
Since it can be shown that the
proportional constant becomes zero if we multiply $N$ to the right,
we get the formula (\ref{uf}).

Now we show the anti-self-duality of $F_{\mu\nu}$.
The field strength of the gauge field (\ref{gauge filed}) is
\begin{eqnarray}
&&F_{\mu\nu}=\partial_{\mu}A_{\nu}-\partial_{\nu}A_{\mu}
-i\,[A_{\mu},\, A_{\nu}]\nonumber\\
&&\hspace{4ex}=i\partial_{\mu}(N^{\dagger}\partial_{\nu}N)
+i(N^{\dagger}\partial_{\mu}N)(N^{\dagger}\partial_{\nu}N)
-(\mu \leftrightarrow\nu)\\
&&\hspace{4ex}=i\partial_{\mu}N^{\dagger}\{I-NN^{\dagger}\}\partial_{\nu}N
-(\mu \leftrightarrow\nu)\nonumber.
\end{eqnarray}
Using the formula (\ref{uf}), we find
\begin{eqnarray}
&&=i(\partial_{\mu}N^{\dagger})MR^{-1}M^{\dagger}(\partial_{\nu}N)
-(\mu \leftrightarrow\nu)\nonumber\\
&&=iN^{\dagger}(\partial_{\mu}M)R^{-1}
(\partial_{\nu}M^{\dagger})N
-(\mu \leftrightarrow\nu)\\
&&=iN^{\dagger}C\sigma_{\mu}R^{-1}\bar{\sigma}_{\nu}C^{T}N
-(\mu \leftrightarrow\nu).\nonumber
\end{eqnarray}
Since $R^{-1}$ is real, it commute with $\sigma_{\mu}$, then we obtain
finally,
\begin{eqnarray}
&&F_{\mu\nu}=iN^{\dagger}CR^{-1}
(\sigma_{\mu}\bar{\sigma}_{\nu}-\sigma_{\nu}\bar{\sigma}_{\mu})C^{T}N
\nonumber\\
&&\hspace{4ex}=-4N^{\dagger}CR^{-1}\bar{\sigma}_{\mu\nu}C^{T}N.
\end{eqnarray}
Because of the anti-self-duality of $\bar{\sigma}_{\mu\nu}$, it is
proved $F_{\mu\nu}$ satisfies the anti-self-dual equation.

\subsubsection{spin 1/2}

The adjoint fermions are given by the zero modes:
\begin{eqnarray}
\Dbarslash\lambda=0,\quad
\Dbarslash\psi=0.
\label{adzeroeq}
\end{eqnarray}
The solutions of these equations have following forms \cite{CGT},
\begin{eqnarray}
&&\lambda^{\dot{r}}_{\alpha\dot{s}}=N^{\dagger\dot{r}r}\left\{{\cal
    M}_r R^{-1}C^T\delta_{\alpha}^s + \epsilon_{r \alpha}CR^{-1}({\cal
    M}^T)^s\right\}N_{s\dot{s}}\nonumber,\\
&&\psi^{\dot{r}}_{\alpha\dot{s}}=N^{\dagger\dot{r}r}\left\{{\cal
    N}_r R^{-1}C^T\delta_{\alpha}^s + \epsilon_{r \alpha}CR^{-1}({\cal
    N}^T)^s\right\}N_{s\dot{s}} .
\label{adjointzero}
\end{eqnarray}
Here ${\cal M}_r$ and ${\cal N}_r$ are constant $(k+1)\times k$ grassmannian
matrices.
These forms of the zero modes are anticipated by the supersymmetry
since there exist zero modes corresponding to the supersymmetry in the
super partner of the gauge field.
If we apply the supersymmetric transformation $\xi$ to $\lambda$, we obtain
\begin{eqnarray}
\delta\lambda= -\bar{\sigma}_{\mu\nu}\xi F_{\mu\nu}
=N^{\dagger\dot{r}r}\left\{4\xi_{r}CR^{-1}C^{T}\delta_{\alpha}^{s}
+\epsilon_{r\alpha}CR^{-1}4\xi^{s}C^{T}\right\}N_{s\dot{s}}.
\end{eqnarray}
The last equation is the same as the equation (\ref{adjointzero}) if we
replace ${\cal M}_r$ with $4\xi_r C$.
To hold (\ref{adzeroeq}), ${\cal M}$ and ${\cal N}$ must
 satisfy the algebraic equations.
Using the formula (\ref{uf}), it can be found
\begin{eqnarray}
\bar{\sigma}_{\mu}^{\dot{\alpha}{\alpha}}
(D_{\mu}\lambda_{\alpha})^{\dot{r}}_{\dot{s}}
=-2iN^{\dagger\dot{r}\alpha}CR^{-1}\left\{ M^{\dagger\alpha s}
{\cal M}_{s}+({\cal M}^{T})^{s}
M_{s\dot{q}}\epsilon^{\dot{\alpha}\dot{q}}\right\}
R^{-1}C^{T}N_{\alpha\dot{s}}.
\end{eqnarray}
Therefore we obtain
\begin{eqnarray}
M^{\dagger\dot{\alpha}s}{\cal M}_s
+({\cal M}^{T})^{s}M_{s \dot{q}}\epsilon^{\dot{\alpha}\dot{q}}=0.
\end{eqnarray}
In the similar way, we find a constraint equation of ${\cal N}$.
\begin{eqnarray}
M^{\dagger\dot{\alpha}s}{\cal N}_s
+({\cal N}^{T})^{s}M_{s \dot{q}}\epsilon^{\dot{\alpha}\dot{q}}=0.
\end{eqnarray}

\subsubsection{spin 0}
The adjoint scalar field satisfies the following equation,
\begin{eqnarray}
D^2\phi-\sqrt{2}i[\lambda,\psi]=0.
\label{theq2}
\end{eqnarray}
Following \cite{DKM}, we first consider the equation without the
source term,
\begin{eqnarray}
D^2\phi_0=0,
\label{th}
\end{eqnarray}
which obeys the boundary condition as $\phi\rightarrow\VEV{\phi}$ at
$x\rightarrow\infty$.
The solution of this equation has the following form:
\begin{eqnarray}
\phi^{\,\dot{r}}_{0\,\dot{s}}=-iN^{\dagger\dot{r}r}A_{r}^{s}N_{s\dot{s}},
\end{eqnarray}
where $A_{r}^{s}$ is the following $(k+1)\times k$ matrix:
\begin{eqnarray}
A_{r}^{s}=\left(
\begin{array}{cccc}
A_{0,0\, r}^{\quad\, s}&0&\cdots&0\\
0&&&\\
\vdots&&\hat{A}\delta_{r}^{s}&\\
0&&&
\end{array}
\right),
\quad A_{0,0}=i\VEV{\phi},\quad\hat{A}^{T}=-\hat{A}.
\end{eqnarray}
Since we take the singular gauge: $N_0\rightarrow 1$,
$N_{0+i}\rightarrow 0$ at
$x\rightarrow\infty$,
$\phi_0$ goes to $\VEV{\phi}$ at $x\rightarrow\infty$.
Substituting this to the left-hand side of (\ref{th}), we obtain
\begin{eqnarray}
&&(D^2\phi_0)^{\dot{r}}_{\dot{s}}
=4iN^{\dagger\dot{r}r}CR^{-1}C^{T}A_{r}^{s}N_{s\dot{s}}
+4iN^{\dagger\dot{r}r}A_{r}^{s}CR^{-1}C^{T}N_{s\dot{s}}\nonumber\\
&&\hspace{9.5ex}
-4iN^{\dagger\dot{r}r}CR^{-1}M^{\dagger\dot{\alpha}\beta}
A_{\beta}^{\gamma}M_{\gamma\dot{\alpha}}R^{-1}C^{T}N_{r\dot{s}}
\nonumber\\
&&\hspace{8.5ex}=4i(N^{\dagger}_{0+i})^{\dot{r}r}R^{-1}_{i,j}
\hat{A}_{j,h}(N_{0+h})_{r\dot{s}}
+4i(N^{\dagger}_{0+i})^{\dot{r}r}\hat{A}_{i,j}R^{-1}_{j,h}(N_{0+h})_{r\dot{s}}
\nonumber\\
&&\hspace{9.5ex}
-4i(N^{\dagger}_{0+i})^{\dot{r}r}R^{-1}_{i,j}{\rm tr}\left(
M^{\dagger}_{j,\,0+m}
\hat{A}_{m,n}M_{n,\,0+l}\right)
R^{-1}_{l,h}(N_{0+h})_{r\dot{s}}\nonumber\\
&&\hspace{9.5ex}
-4i(N^{\dagger}_{0+i})^{\dot{r}r}R^{-1}_{i,j}{\rm tr}\left(
M^{\dagger}_{j,\,0}A_{0,0}M_{0,\,l}\right)
R^{-1}_{l,h}(N_{0+h})_{r\dot{s}}\nonumber\\
&&\hspace{8.5ex}
=4i(N^{\dagger}_{0+i})^{\dot{r}r}\biggl[R^{-1}_{i,j}
\hat{A}_{j,h}
+\hat{A}_{i,j}R^{-1}_{j,h}
-R^{-1}_{i,j}{\rm tr}\left(
M^{\dagger}_{j,\,0+m}
\hat{A}_{m,n}M_{n,\,0+l}\right)
R^{-1}_{l,h}\biggr](N_{0+h})_{r\dot{s}}\nonumber\\
&&\hspace{9.5ex}
-4i(N^{\dagger}_{0+i})^{\dot{r}r}R^{-1}_{i,j}\Lambda_{j,l}
R^{-1}_{l,h}(N_{0+h})_{r\dot{s}},
\label{henkeith}
\end{eqnarray}
where $\Lambda$ is $k\times k$ anti-symmetric matrix defined by
\begin{eqnarray}
&&\Lambda_{j,l}
\equiv{\rm tr}\left(M^{\dagger}_{j,\,0}A_{0,0}M_{0,\,l}\right)\nonumber\\
&&\hspace{4ex}
={\rm tr}(\bar{\omega}_jA_{0,0}\omega_l)\nonumber\\
&&\hspace{4ex}
=\frac{1}{2}{\rm tr}
(A_{0,0}\{\omega_{l}\bar{\omega}_j+\omega_{j}\bar{\omega}_l\})
+\frac{1}{2}{\rm tr}
(A_{0,0}\{\omega_{l}\bar{\omega}_{j}-\omega_{j}\bar{\omega}_{l}\})
\nonumber\\
&&\hspace{4ex}
=\frac{1}{2}{\rm tr}(\bar{\omega}_jA_{0,0}\omega_l
-\bar{\omega}_lA_{0,0}\omega_j).
\end{eqnarray}
Here we have used
$\omega_{l}\bar{\omega}_j+\omega_j\bar{\omega}_l\propto 1$ and
${\rm tr}A_{0,0}=0.$\footnote{For the definition of $\omega$ and
  $\hat{a}$, see (\ref{B}).}
The first term of the right-hand side of (\ref{henkeith}) is
simplified by the following equation,
\begin{eqnarray}
&&R^{-1}_{i,j}\hat{A}_{j,h}+\hat{A}_{i,j}R^{-1}_{j,h}
-R^{-1}_{i,j}{\rm tr}\left(M^{\dagger}_{j,\,0+m}
\hat{A}_{m,n}M_{n,\,0+l}\right)
R^{-1}_{l,h}\nonumber\\
&&=R^{-1}_{i,j}{\rm tr}\left(\frac{1}{2}\hat{A}_{j,m}R_{m,l}
+\frac{1}{2}R_{j,m}\hat{A}_{m,l}-M^{\dagger}_{j,\,0+m}
\hat{A}_{m,n}M_{n,\,0+l}
\right)R^{-1}_{l,h}\nonumber\\
&&=R^{-1}_{i,j}
\left(\left[\hat{a}_{\mu},\left[\hat{a}_{\mu},\hat{A}\right]\,\right]
+\frac{1}{2}\left\{\hat{A},W\right\}\right)_{j,l}R^{-1}_{l,h},
\end{eqnarray}
where $\hat{a}_{\mu}$ is a $k\times k$ real matrix defined by
$\hat{a}=-i\hat{a}_{\mu}\sigma_{\mu}$
and $W$ is a $k\times k$ real symmetric matrix defined by
\begin{eqnarray}
W_{j,l}
=\frac{1}{2}{\rm tr}(\bar{\omega}_{j}\omega_{l}+\bar{\omega}_{l}\omega_{j}).
\end{eqnarray}
Finally we obtain
\begin{eqnarray}
(D^2\phi_0)^{\dot{r}}_{\dot{s}}
=4i(N^{\dagger}_{0+i})^{\dot{r}r}R^{-1}_{i,j}
\left(\left[\hat{a}_{\mu},\left[\hat{a}_{\mu},\hat{A}\right]\,\right]
+\frac{1}{2}\left\{\hat{A},W\right\}-\Lambda\right)_{j,l}R^{-1}_{l,h}
(N_{0+h})_{r\dot{s}}.
\end{eqnarray}
Therefore, $\hat{A}$ must satisfy an algebraic equation,
\begin{eqnarray}
\left[\,\hat{a}_{\mu},\left[\,\hat{a}_{\mu},\hat{A}\,\right]\,\right]
+\frac{1}{2}\left\{\hat{A},W\right\}
=\Lambda.
\label{defhata}
\end{eqnarray}

Now let us solve Eq.(\ref{theq2}).
The solution of (\ref{theq2}) with the boundary condition
$\phi\rightarrow 0$ at $x\rightarrow\infty$ has the following form:
\begin{eqnarray}
\label{phif}
\phi^{\,\dot{r}}_{f\,\dot{s}}=\frac{\sqrt{2}i}{4}
N^{\dagger\dot{r}r}\left\{{\cal N}_rR^{-1}({\cal M}^{T})^s-
{\cal M}_rR^{-1}({\cal N}^{T})^s\right\}N_{s\dot{s}}
-iN^{\dagger\dot{r}r}FN_{r\dot{s}},
\end{eqnarray}
where $F$ is a $(k+1)\times (k+1)$ anti-symmetric matrix:
\begin{eqnarray}
F=\left(
\begin{array}{ccc}
0&\cdots&0\\
\vdots&\hat{F}&\\
0&&
\end{array}
\right)
,\quad
\hat{F}^{T}=-\hat{F}.
\end{eqnarray}
The first term is anticipated by the supersymmetry,
\begin{eqnarray}
\delta\phi=-\sqrt{2}i\xi\psi
=\frac{\sqrt{2}i}{4}N^{\dagger\dot{r}r}
\left\{{\cal N}_rR^{-1}(4\xi^sC^{T})
-(4\xi_rC)R^{-1}({\cal N}^{T})^s\right\}
N_{s\dot{s}}.
\end{eqnarray}
Due to the existence of the zero modes which are not related with
supersymmetry, the second term of Eq.(\ref{phif}) is necessary.
By a tedious but straightforward calculation, it is found
\begin{eqnarray}
&&(D^2\phi_f)^{\dot{r}}_{\dot{s}}-i\sqrt{2}[\lambda,\psi]^{\dot{r}}_{\dot{s}}
=4iN^{\dagger\dot{r}r}CR^{-1}C^{T}FN_{r\dot{s}}
+4iN^{\dagger\dot{r}r}FCR^{-1}C^{T}N_{r\dot{s}}\nonumber\\
&&\hspace{22ex}
-4iN^{\dagger\dot{r}r}CR^{-1}
M^{\dagger\dot{\alpha}\beta}FM_{\beta\dot{\alpha}}
R^{-1}C^{T}N_{r\dot{s}}
\nonumber\\
&&\hspace{22ex}+\sqrt{2}iN^{\dagger\dot{r}r}CR^{-1}
\left\{({\cal M}^{T})^s{\cal N}_s-({\cal N}^{T})^s{\cal M}_s\right\}
R^{-1}C^{T}N_{r\dot{s}}\nonumber\\
&&\hspace{21.3ex}
=4i(N^{\dagger}_{0+i})^{\dot{r}r}\biggl[R^{-1}_{i,j}
\hat{F}_{j,h}
+\hat{F}_{i,j}R^{-1}_{j,h}
-R^{-1}_{i,j}{\rm tr}\left(
M^{\dagger}_{j,\,0+m}
\hat{F}_{m,n}M_{n,\,0+l}\right)
R^{-1}_{l,h}\biggr](N_{0+h})_{r\dot{s}}\nonumber\\
&&\hspace{22ex}
+\sqrt{2}i(N^{\dagger}_{0+i})^{\dot{r}r}R^{-1}_{i,j}
\left\{({\cal M}^{T})^s{\cal N}_s-({\cal N}^{T})^s{\cal M}_s\right\}_{j,l}
R^{-1}_{l,h}(N_{0+h})_{r\dot{s}}.
\label{henkeithf}
\end{eqnarray}
The right-hand side of the above equation is the same
as the right-hand side of (\ref{henkeith}) if the following
replacement is performed:
\begin{eqnarray}
\hat{F}\rightarrow\hat{A},\quad
-\frac{\sqrt{2}}{4}
\left\{({\cal M}^{T})^s{\cal N}_s-({\cal N}^{T})^{s}{\cal M}_s\right\}
\rightarrow\Lambda.
\end{eqnarray}
Thus as well as $\hat{A}$, the following algebraic equation is derived,
\begin{eqnarray}
\left[\,\hat{a}_{\mu},\left[\,\hat{a}_{\mu},\hat{F}\,\right]\,\right]
+\frac{1}{2}\left\{\hat{F},W\right\}
=-\frac{\sqrt{2}}{4}
\left\{({\cal M}^{T})^s{\cal N}_s-({\cal N}^{T})^{s}{\cal
    M}_s\right\}.
\label{defhatf}
\end{eqnarray}
To satisfy the boundary condition $\phi\rightarrow\VEV{\phi}$ at
$x\rightarrow\infty$, we add $\phi_0$ to $\phi_f$,
\begin{eqnarray}
\phi=\phi_0+\phi_f.
\end{eqnarray}
This is the solution of (\ref{theq2}) \cite{DKM}.

In $N=2$ supersymmetric QCD, there appears a source term in the equation
of the adjoint anti-scalar field,
\begin{eqnarray}
D^2\phi^{\dagger a}-\sqrt{2}i\tilde{q}T^a q=0.
\end{eqnarray}
The source term comes from the Yukawa term $\sqrt{2}i\tilde{q}\phi q$.
The fundamental zero modes $q,\tilde{q}$ will be explained in detail in the
next subsection.
{}From the completeness condition of $T^a$,
\begin{eqnarray}
(T^a)^{\dot{u}}_{\dot{t}}(T^a)^{\dot{r}}_{\dot{s}}
=\frac{1}{2}\delta^{\dot{u}}_{\dot{s}}\delta^{\dot{r}}_{\dot{t}}
-\frac{1}{4}\delta^{\dot{u}}_{\dot{t}}\delta^{\dot{r}}_{\dot{s}},
\end{eqnarray}
the above equation becomes
\begin{eqnarray}
(D^2\phi^{\dagger})^{\dot{r}}_{\dot{s}}-
\frac{\sqrt{2}i}{4}\left\{2\tilde{q}^{\alpha}_{\dot{s}}q^{\dot{r}}_{\alpha}
-\delta^{\dot{r}}_{\dot{s}}
\tilde{q}^{\alpha}_{\dot{t}}q^{\dot{t}}_{\alpha}\right\}=0.
\label{eqphidagger}
\end{eqnarray}
Substituting the fundamental zero modes given in the next subsection,
the source term becomes
\begin{eqnarray}
\frac{\sqrt{2}i}{4}\left\{2\tilde{q}^{\alpha}_{\dot{s}}q^{\dot{r}}_{\alpha}
-\delta^{\dot{r}}_{\dot{s}}
\tilde{q}^{\alpha}_{\dot{t}}q^{\dot{t}}_{\alpha}\right\}
=4iN^{\dagger\dot{r}r}CR^{-1}Z R^{-1}C^{T}N_{r\dot{s}},
\label{qqsource}
\end{eqnarray}
where $Z$ is a $k\times k$ grassmannian matrix given by
\begin{eqnarray}
Z_{h,l}=-\frac{\sqrt{2}}{16\pi^2}
(\zeta_{h}\tilde{\zeta}_{l}-\zeta_{l}\tilde{\zeta}_{h}).
\end{eqnarray}
Eq.(\ref{eqphidagger}) can be solved by the following ansatz:
\begin{eqnarray}
\phi^{\dagger}_q=-iN^{\dagger \dot{r}r}PN_{r \dot{s}},
\end{eqnarray}
where $P$ is a $(k+1)\times (k+1)$ even grassmannian matrix having the
following form:
\begin{eqnarray}
P=\left(
\begin{array}{ccc}
0&\cdots&0\\
\vdots&\hat{P}&\\
0&&
\end{array}
\right),\quad \hat{P}^{T}=-\hat{P}.
\end{eqnarray}
Under this ansatz, we obtain
\begin{eqnarray}
&&(D^2\phi^{\dagger}_q)^{\dot{r}}_{\dot{s}}-
\frac{\sqrt{2}i}{4}\left\{2\tilde{q}^{\alpha}_{\dot{s}}q^{\dot{r}}_{\alpha}
-\delta^{\dot{r}}_{\dot{s}}
\tilde{q}^{\alpha}_{\dot{t}}q^{\dot{t}}_{\alpha}\right\}\nonumber\\
&&=4iN^{\dagger\dot{r}r}CR^{-1}C^{T}PN_{r\dot{s}}
+4iN^{\dagger\dot{r}r}PCR^{-1}C^{T}N_{r\dot{s}}\nonumber\\
&&\hspace{1ex}
-4iN^{\dagger\dot{r}r}CR^{-1}
M^{\dagger\dot{\alpha}\beta}PM_{\beta\dot{\alpha}}
R^{-1}C^{T}N_{r\dot{s}}
-4iN^{\dagger\dot{r}r}CR^{-1}Z R^{-1}C^{T}N_{r\dot{s}}
\nonumber\\
&&=4i(N^{\dagger}_{0+i})^{\dot{r}r}\biggl[R^{-1}_{i,j}
\hat{P}_{j,h}
+\hat{P}_{i,j}R^{-1}_{j,h}
-R^{-1}_{i,j}{\rm tr}\left(
M^{\dagger}_{j,\,0+m}
\hat{P}_{m,n}M_{n,\,0+l}\right)
R^{-1}_{l,h}\biggr](N_{0+h})_{r\dot{s}}\nonumber\\
&&\hspace{1ex}-4i(N^{\dagger}_{0+i})^{\dot{r}r}R^{-1}_{i,j}
Z_{j,l}R^{-1}_{l,h}(N_{0+h})_{r\dot{s}}.
\end{eqnarray}
As well as Eq.(\ref{henkeithf}), the right-hand side of the above
equation is the same as the right-hand side of Eq.(\ref{henkeith})
if we replace $\hat{P}$ and $Z$ with $\hat{A}$ and $\Lambda$.
Therefore, from Eq.(\ref{eqphidagger}) the matrix $\hat{P}$ must
satisfy the following algebraic equation:
\begin{eqnarray}
\left[\,\hat{a}_{\mu},\left[\,\hat{a}_{\mu},\hat{P}\,\right]\,\right]
+\frac{1}{2}\left\{\hat{P},W\right\}
=Z.
\label{defhatp}
\end{eqnarray}
The above solution $\phi_q^{\dagger}$ does not satisfy the boundary condition
$\phi^{\dagger}\rightarrow\VEV{\phi}^{\dagger}$ at $x\rightarrow\infty$.
To satisfy this boundary condition, we add $\phi_0^{\dagger}$ to the
above solution.
Thus we obtain
\begin{eqnarray}
\phi^{\dagger}=\phi_0^{\dagger}+\phi_q^{\dagger}.
\end{eqnarray}

\subsection{hypermultiplet}

\subsubsection{spin 1/2}
As is known by the index theorem, for $k$-instanton there exist $k$ zero
modes in the fundamental fermion $q$ and $\tilde{q}$ respectively.
The explicit forms of the zero modes
\begin{eqnarray}
\Dbarslash q=0,\quad
\Dbarslash \tilde{q}=0
\end{eqnarray}
are given by \cite{OCFGT}
\begin{eqnarray}
  \label{SQCDsol}
q_{f\alpha} ^{\dot{r}}=-\Psi^{\dagger\dot{r}r}\epsilon_{r\alpha}\zeta_f,
\quad
\tilde{q}^{\alpha}_{f\dot{s}}
=-\tilde{\zeta}_{f}^{T}\epsilon^{\alpha s}\Psi_{s\dot{s}},
\end{eqnarray}
where $\Psi$ is a $k$-dimensional quaternion column vector given by
\begin{eqnarray}
\label{zeromode}
\Psi_{s\dot{s}}=\frac{1}{\pi}R^{-1}C^{T}N_{s\dot{s}},\quad
\Psi^{\dagger\dot{r}r}=\frac{1}{\pi}N^{\dagger\dot{r}r}CR^{-1},
\end{eqnarray}
and $\zeta_f$ and $\tilde{\zeta}_f$ are $k$-dimensional grassmannian column
vectors which exactly correspond to $k$ zero modes in $q$ and $\tilde{q}$.
The index $f=1,\cdots N_f$ is the flavor one.
$\Psi$ is normalized as
\begin{eqnarray}
\int d^4x{\rm tr}(\Psi_i\bar{\Psi}_j)=\delta_{i,j},
\label{normalization}
\end{eqnarray}
which is proved by an identity
\begin{eqnarray}
\partial_{\mu}^2 R^{-1}=-4R^{-1}C^{T}{\rm tr}(NN^{\dagger})CR^{-1},
\end{eqnarray}
and $R^{-1}_{i,j}\rightarrow\delta_{i,j}/x^2$ at $x\rightarrow\infty$.
\subsubsection{spin 0}
The defining equations of the fundamental scalar fields are
\begin{eqnarray}
D^2Q-\sqrt{2}i\lambda q=0,\quad
D^2\tilde{Q}+\sqrt{2}i\tilde{q}\lambda=0,\\
D^2Q^{\dagger}-\sqrt{2}i\tilde{q}\psi=0,\quad
D^2\tilde{Q}^{\dagger}-\sqrt{2}i\psi q=0.
\end{eqnarray}
The supersymmetry gives a clue to solve these equations.
Under the supersymmetry $\xi$, $Q$ transforms as
\begin{eqnarray}
\delta Q=-\sqrt{2}i\xi q
=\frac{\sqrt{2}i}{4}N^{\hat{r}r}(4\xi_rC)R^{-1}\zeta_f.
\end{eqnarray}
Thus as well as $\lambda$, we anticipate that the solution of the
above equation is obtained by $4\xi_r C\rightarrow {\cal M}_r$.
This is indeed the case and using the $SU(2)_R$ symmetry, we find all the
solutions of the defining equations:
\begin{eqnarray}
&&Q^{\dot{r}}_f=\frac{\sqrt{2}i}{4}N^{\dagger \dot{r}r}
{\cal M}_rR^{-1}\zeta_f,\quad
\tilde{Q}_{f\dot{r}}=-\frac{\sqrt{2}i}{4}
\epsilon_{\dot{r}\dot{s}}N^{\dagger \dot{s}r}{\cal M}_rR^{-1}
\tilde{\zeta}_f,\\
&&Q^{\dagger}_{f\dot{r}}=\frac{\sqrt{2}i}{4}
\epsilon_{\dot{r}\dot{s}}N^{\dagger \dot{s}r}{\cal N}_rR^{-1}
\tilde{\zeta}_f,\quad
\tilde{Q}^{\dagger\dot{r}}_f=\frac{\sqrt{2}i}{4}
N^{\dagger \dot{r}r}{\cal N}_rR^{-1}\zeta_f.
\end{eqnarray}
The $SU(2)_R$ symmetry is manifest in the above solutions.
\subsection{supersymmetric instanton action}
Let us compute the action of the supersymmetric instanton in $N=2$
supersymmetric QCD.
As was shown in section \ref{sec-def}, the leading part of the
Lagrangian is
\begin{eqnarray}
&&{\cal L}_0=
\frac{1}{2}{\rm tr}(F_{\mu\nu}F_{\mu\nu})\nonumber\\
&&\hspace{3ex}+{\rm tr}\left\{-2i\bar{\lambda}\Dbarslash\lambda
-2i\bar{\psi}\Dbarslash\psi+2(D_{\mu}\phi)^{\dagger}D_{\mu}\phi
+2\sqrt{2}i\lambda[\psi,\phi^{\dagger}]\right\}
\nonumber\\
&&\hspace{3ex}+(D_{\mu}Q)^{\dagger}D_{\mu}Q
+D_{\mu}\tilde{Q}(D_{\mu}\tilde{Q})^{\dagger}
 -i\bar{q}\Dbarslash q -i\bar{\tilde{q}}\Dbarslash\tilde{q}\nonumber\\
&&\hspace{3ex}+\sqrt{2}i\left(\tilde{q}\phi q+Q^{\dagger}\lambda
  q-\tilde{q}\lambda\tilde{Q}^{\dagger}
+\tilde{q}\psi Q+\tilde{Q}\psi q\right).
\label{lagmatter}
\end{eqnarray}
Substituting the solution of the defining equations for ${\cal L}_0$
and integrating it, the supersymmetric instanton action $S_0$ is
obtained.
To carry out the integration, we convert the volume integration to the
surface integration by the equation of supersymmetric instanton.
{}From the defining equations, we find
\begin{eqnarray}
S_0=\frac{8k\pi^2}{g^2}
+\int d^4x\partial_{\mu}\left\{{\rm tr}(2\phi^{\dagger}D_{\mu}\phi)
+(D_{\mu}Q)^{\dagger}Q+(D_{\mu}\tilde{Q})^{\dagger}\tilde{Q}\right\}
+\int d^4x \sqrt{2}i\left(\tilde{q}\phi q+Q^{\dagger}\lambda q
+\tilde{Q}\psi q\right).\hspace{-6ex}\nonumber\\
\label{leadingaction1}
\end{eqnarray}
To integrate the last term, we introduce the auxiliary solution $\bar{q}$,
\begin{eqnarray}
  \label{qbar}
  \bar{q}^{\dot{\alpha}}_{f\dot{r}}=\frac{1}{4\pi}
\tilde{\zeta}^{T}R^{-1}M^{\dagger\dot{\alpha}s}
\left\{{\cal M}_sR^{-1}({\cal N}^T)^r-{\cal N}_sR^{-1}({\cal M}^T)^r
\right\}N_{r\dot{r}}
+\frac{1}{\sqrt{2}}\tilde{\zeta}^TR^{-1}M^{\dagger\dot{\alpha}r}FN_{r\dot{r}},
\end{eqnarray}
which satisfies the equation:
\begin{eqnarray}
\label{eqqbar}
\Dslash \bar{q}
+\sqrt{2}Q^{\dagger}\lambda+\sqrt{2}\tilde{q}\phi_f
+\sqrt{2}\tilde{Q}\psi
=i\sqrt{2}\tilde{\zeta}^{T}\hat{F}\Psi.
\end{eqnarray}
Using the auxiliary solution $\bar{q}$ and $\phi_q^{\dagger}$, the
last term of Eq.(\ref{leadingaction1}) becomes
\begin{eqnarray}
&&\sqrt{2}i(\tilde{q}\phi q+Q^{\dagger}\lambda q+\tilde{Q}\psi q)
\nonumber\\
&&\hspace{3ex}={\rm tr}\left\{2(D^2\phi^{\dagger})\phi_0\right\}
-i(\Dslash\bar{q})^{\alpha}_{\dot{r}}q_{\alpha}^{\dot{r}}
+\sqrt{2}\sum_{f=1}^{N_f}\tilde{\zeta}^{T}_f\hat{F}
\Psi_{\alpha\dot{r}}q^{\alpha\dot{r}}_f\nonumber\\
&&\hspace{3ex}=\partial_{\mu}
{\rm tr}\left\{2(D_{\mu}\phi^{\dagger})\phi_0
-2\phi^{\dagger}(D_{\mu}\phi_0)\right\}
-i\partial_{\mu}(\bar{q}\sigma_{\mu}q)
+\sqrt{2}\sum_{f=1}^{N_f}\tilde{\zeta}^{T}_f\hat{F}
{\rm tr}(\Psi\Psi^{\dagger})\zeta_f.
\end{eqnarray}
Therefore, the leading action is
\begin{eqnarray}
&&S_0=\frac{8k\pi^2}{g^2}
+\int d^4x\partial_{\mu}{\rm tr}\left\{2\phi^{\dagger}D_{\mu}\phi
+2(D_{\mu}\phi)^{\dagger}\phi_0
-2\phi^{\dagger}D_{\mu}\phi_0\right\}\nonumber\\
&&\hspace{5ex}+\int d^4x\partial_{\mu}\left\{
(D_{\mu}Q)^{\dagger}Q+(D_{\mu}\tilde{Q})^{\dagger}\tilde{Q}
-i\bar{q}\sigma_{\mu}q\right\}
+\sqrt{2}\sum_{f=1}^{N_f}\tilde{\zeta}^{T}_f\hat{F}\zeta_f,
\end{eqnarray}
where we have used the normalization condition of $\Psi$.
Now all the volume integrations reduce to the surface integrations.
The surface integrations are evaluated by the asymptotic behavior of the
supersymmetric instanton.
Since at infinity $M$, $R^{-1}$ and $N$ behaves as
\begin{eqnarray}
&&M\rightarrow -Cx,\quad R^{-1}\rightarrow\frac{1}{|x|^2},\nonumber\\
&&N_0\rightarrow 1-\frac{|\omega|_i^2}{2|x|^2},\quad
N_{0+i}\rightarrow\frac{x\bar{\omega_i}}{|x|^2},
\end{eqnarray}
we have the following asymptotic behaviors of the supersymmetric instanton:
\begin{eqnarray}
&&A_{\mu\dot{s}}^{\dot{r}}\rightarrow
\frac{2}{x^4}\bar{\sigma}_{0}^{\dot{r}r}(\omega_i)_{r\dot{\alpha}}
(\sigma_{\mu\nu})^{\dot{\alpha}}_{\dot{\beta}}
(\bar{\omega}_i)^{\dot{\beta}s}\sigma_{0s\dot{s}}x_{\nu},\nonumber\\
&&\phi_{0\dot{s}}^{\dot{r}}\rightarrow
-i\bar{\sigma}_0^{\dot{r}r}\left\{(A_{0,0})_{r}^{s}
\left(1-\frac{|\omega_i|^2}{x^2}\right)
+\frac{1}{x^2}(\omega_{i})_{r\dot{t}}\hat{A}_{i,j}(\bar{\omega}_j)^{\dot{t}s}
\right\}\sigma_{0s\dot{s}},\nonumber\\
&&\phi_{f\dot{s}}^{\dot{r}}\rightarrow-i\bar{\sigma}_0^{\dot{r}r}\left\{
-\frac{\sqrt{2}}{4x^2}
\left\{({\cal N}_{0,i})_r({\cal M}_{0,i})^s
-({\cal M}_{0,i})_r({\cal N}_{0,i})^s\right\}
+\frac{1}{x^2}(\omega_{i})_{r\dot{t}}\hat{F}_{i,j}(\bar{\omega}_j)^{\dot{t}s}
\right\}\sigma_{0s\dot{s}},\nonumber\\
&&\phi_{q\dot{s}}^{\dagger\dot{r}}\rightarrow
-i\bar{\sigma}_0^{\dot{r}r}\left\{
\frac{1}{x^2}(\omega_{i})_{r\dot{t}}\hat{P}_{i,j}(\bar{\omega}_j)^{\dot{t}s}
\right\}\sigma_{0s\dot{s}}.
\end{eqnarray}
We have omitted the asymptotic behaviors of the hypermultiplets, since
they vanish faster than $O(x^{-2})$ and do not contribute the integration.
Finally we obtain the following supersymmetric instanton action:
\begin{eqnarray}
&&S_0=\frac{8k\pi^2}{g^2}+S_{higgs}+S_{yukawa},\nonumber\\
&&S_{higgs}=16\pi^2|A_{0,0}|^2|\omega_i|^2
-8\pi^2{\rm tr}(\bar{\omega}_iA^{\dagger}_{0,0}\omega_j)\hat{A}_{j,i}
\nonumber\\
&&S_{yukawa}=2\sqrt{2}\pi^2
{\rm tr}\left\{
{\cal N}_{0,i}A^{\dagger}_{0,0}{\cal M}_{0,i}
-{\cal M}_{0,i}A^{\dagger}_{0,0}{\cal N}_{0,i}
\right\}
-8\pi^2{\rm tr}(\bar{\omega}_iA^{\dagger}_{0,0}\omega_j)\hat{F}_{j,i}
\nonumber\\
&&\hspace{10ex}
+8\pi^2\hat{P}_{i,j}
{\rm tr}(\bar{\omega}_jA_{0,0}\omega_i)
+\sqrt{2}\sum_{f=1}^{N_f}\tilde{\zeta}^{T}_f\hat{F}\zeta_f,
\end{eqnarray}
where $|A_{0,0}|^2={\rm tr}\{A_{0,0}^{\dagger}A_{0,0}\}/2$.

\reseteqnum
\section{Instanton calculus}
In this section, we will perform the instanton calculus
by using the supersymmetric instanton constructed in the
previous sections.
We calculate the moduli $u=\VEV{{\rm tr}\phi^2}$ as the function of
$\VEV{\phi}$. We choose $\VEV{\phi}=a \sigma_3/2 $, where $a$ is
real. In the classical limit, $u\sim a^2/2$ in this convention, which
differs from $a$ in section 2 by factor 2.
Taking into account the super transformations, it is easy to find that
the adjoint scalar $\phi$ contains the following part;
\begin{eqnarray}
  \label{superphi}
  \phi=-\sqrt{2}i\xi\psi+\cdots
=\sqrt{2}i\xi\bar{\sigma}_{\mu\nu}\xi'F_{\mu\nu}+\cdots,
\end{eqnarray}
where $\xi$, $\xi'$ are the grassmannian collective coordinates
accompanying $N=2$ supersymmetry
and
$\cdots$ includes the other fermionic zero modes and $\phi_0$.
Then, ${\rm tr}\phi^2$ is given by,
\begin{eqnarray}
&& {\rm tr}\phi^2 =-2{\rm
    tr}\left[\left(\xi\bar{\sigma}_{\mu\nu}\xi'F_{\mu\nu}\right)^2\right]
+\cdots\\
&&\hspace{2ex}=-\xi^2\xi'^2{\rm
tr}\left(F_{\mu\nu}F_{\mu\nu}\right)+\cdots\, .
\end{eqnarray}
Therefore supersymmetric zero modes are saturated by inserting ${\rm
  tr}\phi^2$, and
we obtain the following result by performing the integration over the center
of the instanton;
\begin{eqnarray}
  \label{ssmode}
  \int d^4 x_0\int d^2\xi d^2\xi'{\rm tr}\phi^2=-\int d^4 x_0 {\rm
    tr}\left[F_{\mu\nu}(x-x_0)F_{\mu\nu}(x-x_0)\right] =-16\pi^2k \, ,
\end{eqnarray}
for $k$-instanton. The other fermionic modes are lifted by the Yukawa
terms in the action and saturated by pulling down those terms from the
action.

For one-instanton sector, the result is given by \cite{FP},
\begin{eqnarray}
\label{oneinst}
  u_1=\frac{1}{2}a^2 \cdot 2 \left(\frac{\Lambda_0}{a}\right)^4,
\end{eqnarray}
for $N_f=0$. We denote $k$-instanton correction to $u$ by $u_k$. For
massless $N_f>0$, $u_1$ vanishes due to the parity symmetry.

For $N_f>0$, the leading-order correction comes from the two-instanton
sector.
First we construct the supersymmetric two-instanton, according to the previous
section. In the two-instanton sector, $\hat{a}$ is a 2$\times$2
matrix made up of quaternions;
\begin{eqnarray}
\label{hata}
\hat{a}=\left(
  \begin{array}{cc}
     x_0+a_3&a_1\\
    a_1&x_0-a_3
  \end{array}
\right), \quad a_1=\frac{a_3}{4|a_3|^2}
\left(\bar{\omega}_2\omega_1-\bar{\omega}_1\omega_2
\right),
\end{eqnarray}
which satisfies the reality condition of $R=M^{\dagger}M$.
We define the grassmannian collective coordinates by the following;
\begin{eqnarray}
  \label{collect}
&&{\cal M}_s=\left(
  \begin{array}{cc}
    \mu_{1s}&\mu_{2s}\\
    4\xi_s+m_{3s}&m_{1s}\\
    m_{1s}&4\xi_s-m_{3s}
  \end{array}
\right),
\quad\quad {\cal N}_s=\left(
  \begin{array}{cc}
    \nu_{1s}&\nu_{2s}\\
    4\xi'_s+n_{3s}&n_{1s}\\
    n_{1s}&4\xi'_s-n_{3s}
  \end{array}
\right),
\\
&&m_1=\frac{a_3}{2|a_3|^2}
\left(2\bar{a}_1m_3+\bar{\omega}_2\mu_1-\bar{\omega}_1\mu_2\right)
,\quad n_1=\frac{a_3}{2|a_3|^2}
\left(2\bar{a}_1n_3+\bar{\omega}_2\nu_1-\bar{\omega}_1\nu_2\right),
\end{eqnarray}
which satisfy the constraint relations for ${\cal M}_s$ and ${\cal
  N}_s$, respectively.
By solving Eq.(\ref{defhata}), Eq.(\ref{defhatf}) and Eq.(\ref{defhatp}), the
matrix $\hat{A}$,
$\hat{F}$ and $\hat{P}$ are given by
\begin{eqnarray}
&&\hat{A}=\left(
\begin{array}{cc}
0&\alpha\\
-\alpha&0
\end{array}
\right),
\quad \hat{F}=\left(
\begin{array}{cc}
0&\beta\\
-\beta&0
\end{array}
\right),
\quad  \hat{P}=\left(
\begin{array}{cc}
0&\gamma\\
-\gamma&0
\end{array}
\right),\\
&&\alpha=-\frac{\omega}{H},\\
&&\beta=-\frac{\sqrt{2}}{4H}\left(\mu_1\nu_2-\mu_2\nu_1+2m_3n_1-2m_1n_3\right),
\\
&&\gamma=\frac{\sqrt{2}}{16\pi^2}
\frac{\sum_{f=1}^{N_f}\tilde{\zeta}_f
\zeta_f}{H},
\end{eqnarray}
where
\begin{eqnarray}
&&L=|\omega_1|^2+|\omega_2|^2,\quad
H=L+4|a_1|^2+4|a_3|^2,\\
\nonumber
&&\Omega=\omega_1\bar{\omega}_2-\omega_2\bar{\omega}_1,\quad
\omega=\frac{1}{2}{\rm tr}\left(\Omega A_{0,0}\right),
\end{eqnarray}
and
$\tilde{\zeta}_f\zeta_f=\tilde{\zeta}^i_f\zeta_{fi}
=\epsilon^{ij}\tilde{\zeta}_{fj}\zeta_{fi}
=\tilde{\zeta}_{f2}\zeta_{f1}-\tilde{\zeta}_{f1}\zeta_{f2}$.
By using the above collective coordinates, the action of
supersymmetric instanton becomes
\begin{eqnarray}
  \label{action}
&&S=\frac{16\pi^2}{g^2}+S_{higgs}+S_{yukawa},\nonumber\\
&&S_{higgs}=16\pi^2\left(L|A_{0,0}|^2-\frac{\omega^2}{H}\right),\\
&&S_{yukawa}=-4\sqrt{2}\pi^2\left\{\nu_kA_{0,0}\mu_k
+\frac{\omega}{H}\left(\mu_1\nu_2-\mu_2\nu_1+2m_3n_1-2m_1n_3\right)
\right\}
\nonumber\\
&&\hspace{8ex}+\frac{1}{2H}\left(
\mu_1\nu_2-\mu_2\nu_1+2m_3n_1-2m_1n_3\right)
\sum_{f=1}^{N_f}\tilde{\zeta}_f\zeta_f
+\sqrt{2}\frac{\omega}{H}\sum_{f=1}^{N_f}\tilde{\zeta}_f\zeta_f,
\nonumber
\end{eqnarray}
where $A_{0,0}=i\VEV{\phi}=ia\sigma^3/2$ and then $|A_{0,0}|^2=a^2/4$.
Compared with the pure Yang-Mills case\cite{DKM}, the last two terms in
$S_{yukawa}$ are added.
Note that a biquadratic term in grassmannian variables appears in the action.
This is a new feature in $N=2$ supersymmetric QCD.
The measure of the collective coordinate is given by
\footnote{The definition of the numerical factor of the measure
  depends on the regularization
  scheme. See appendix D for further details.}
\cite{Osborn,DKM},
\begin{eqnarray}
  \label{measure}
&&C_J\int d^4x_0d^4a_3 d^4\omega_1
d^4\omega_2d^2\xi d^2m_3d^2\mu_1d^2\mu_2d^2\xi'd^2n_3d^2\nu_1d^2\nu_2\\
\nonumber
&&\hspace{5ex}\times
\prod_{f=1}^{N_f}d^2\zeta_f d^2\tilde{\zeta}_f
\frac{\left|\left|a_3\right|^2-\left|a_1\right|^2\right|}{H}
\exp\left(-S_{higgs}-S_{yukawa}\right),\\
&&C_J=2^{6+2 N_f}\pi^{-8}\Lambda_{N_f}^{8-2N_f},
\end{eqnarray}
where $\Lambda_{N_f}^{8-2N_f}$ is replaced by $q=e^{2 \pi i \tau}$
for $N_f=4$.

The supersymmetric zero modes $\xi$ and $\xi'$ are saturated by
inserting ${\rm tr}\phi^2$, as we have already seen.
The other fermionic modes are lifted by the Yukawa terms in the
action, and integrating out those modes except $\zeta_f$,
$\tilde{\zeta}_f$, we obtain
\begin{eqnarray}
  \nonumber
&&\int d^2 m_3 d^2 \mu_1 d^2\mu_2 d^2 n_3 d^2 \nu_1 d^2\nu_2
  \exp\left(-S_{yukawa}\right) \\
\label{yukawa}
&&\hspace{5ex}=-\left(
\frac{16\sqrt{2}\pi^6}{|a_3|^2H|\Omega|}\right)^2
f(y)\exp\left(-\sqrt{2}\frac{\omega}{H}
\sum_{f=1}^{N_f}\tilde{\zeta}_f\zeta_f\right),
\end{eqnarray}
where
\begin{eqnarray}
  \label{f(y)}
 &&f(y)=\omega^2y^2
\left\{
  \left(|\Omega|^2|A_{0,0}|^2+\frac{L\omega^2 y}{H}\right)^2
  +\frac{L^2-|\Omega|^2}{H^2} \omega^2 y^2
\left(|A_{0,0}|^2|\Omega|^2-\omega^2\right)
\right\},\\
  \label{y}
 &&\hspace{5ex}y=1-\frac{\sqrt{2}}{16\pi^2\omega}\sum_{f=1}^{N_f}
\tilde{\zeta}_f\zeta_f.
\end{eqnarray}
The remaining grassmannian integrations are performed as follows;
\begin{eqnarray}
  \nonumber
&& \int \prod_{f=1}^{N_f}d^2\tilde{\zeta}_fd^2\zeta_f
  f(y)\exp\left(-\sqrt{2}\frac{\omega}{H}
\sum_{g=1}^{N_f}\tilde{\zeta}_g\zeta_g\right)
=\left(-\frac{1}{2}
\frac{\omega^2}{H^2}\right)^{N_f}
\sum_{k=0}^{2N_f}\,_{2N_f}\hspace{-0.3ex}C_k
\left(\frac{H}{16\pi^2\omega^2}\right)^k\left.\frac{\partial^k f}
{\partial y^k}\right|_{y=1}.\\
\label{zetaint}
\end{eqnarray}
We change the integration variables from $a_3,\omega_1,\omega_2$ to
$H,L,\Omega$, and then the measure of the integration becomes,
\begin{eqnarray}
  \label{bosonicmeasure}
&&\int d^4 a_3\frac{\left|
      \left|a_3\right|^2-\left|a_1\right|^2\right|}{|a_3|^4}
=\frac{\pi^2}{2}\int_{L+2|\Omega|}^{\infty}dH,\\
&&\int d^4\omega_1
d^4\omega_2=\frac{\pi^3}{8}\int_0^{\infty}dL\int_{|\Omega|\leq
L}d^3\Omega .
\end{eqnarray}
With the change to a polar
coordinate: $\omega=|\Omega||A_{0,0}|\cos\theta$ and
the rescaling: $\Omega'=\Omega /L$ and $H'=H/L$,
the measure is given by,
\begin{eqnarray}
\nonumber
&&\frac{\pi^5}{16}\int_0^{\infty}d L\int_{|\Omega|\leq L}d^3\Omega
  \int_{L+2|\Omega|}^{\infty}dH\\
\label{mrescale}
&&\hspace{3ex}=\frac{\pi^6}{8}\int_0^{\infty}dL L^4\int_{-1}^1d(\cos
\theta)\int_0^1|\Omega'|^2d|\Omega'|\int_{1+2|\Omega'|}^{\infty}dH' ,
\end{eqnarray}
and  $f(y)$ becomes
\begin{eqnarray}
  \label{frescale}
  f(y)=|A_{0,0}|^6|\Omega'|^6L^6\cos^2\theta \, G(y;|\Omega'|,H',\theta),
\end{eqnarray}
where
\begin{eqnarray}
  \label{G(y)}
  G(y;|\Omega'|,H',\theta)=y^2\left\{\left(1+\frac{y}{H'}\cos^2\theta\right)^2
+\frac{1-|\Omega'|^2}{4H'^2}y^2\sin^2 2\theta\right\}.
\end{eqnarray}
Using Eq.(\ref{ssmode}), (\ref{measure}), (\ref{yukawa}), (\ref{zetaint}),
(\ref{mrescale}) and (\ref{frescale}) and performing the integration
of $L$, we obtain the two-instanton
correction to $u$,
\begin{eqnarray}
  \label{vevu}
  u_{2}=\frac{1}{2} a^2\left(\frac{\Lambda_{N_f}}{ a}\right)^{8-2N_f}
  \cdot\left(-\frac{1}{2}\right)^{N_f} I(N_f) ,
\end{eqnarray}
where $I(N_f)$ is defined by
\begin{eqnarray}
  \label{inf}
&&I(N_f)=\int_{-1}^1d(\cos\theta)\cos^2\theta\int_0^1d|\Omega'||\Omega'|^6
\int_{1+2|\Omega'|}^{\infty}\frac{dH'}{H'^3} \left(\frac{|\Omega'|\cos \theta
    }{H'}\right)^{2N_f}\sum_{k=0}^{K}\,_{2N_f}\hspace{-0.3ex}C_k\,
\\
\nonumber
&&\hspace{10ex}\times(5-k)!
\left(1-\frac{|\Omega'|^2\cos^2\theta}{H'}\right)^{k-6}
\left(\frac{H'}{|\Omega'|^2
\cos^2\theta}\right)^k\left.\frac{\partial^k}{\partial  y^k}
G(y;|\Omega'|,H',\theta)\right|_{y=1},
\end{eqnarray}
and $K={\rm min}[4,2N_f]$. The evaluation of  $I(N_f)$ is complicated but
straightforward.
Finally we obtain
\begin{eqnarray}
\label{instcal}
  u_2=\frac{1}{2}a^2 \times \left\{
    \begin{array}{lcc}
\vspace{1ex}
      \displaystyle
\frac{5}{2}\left(\frac{\Lambda_0}{a}\right)^8&{\rm for}&N_f=0\,,\\
\vspace{1ex}
      \displaystyle
(-1)^{N_f}\frac{2N_f-1}{2^{N_f+1} 3^{2
    N_f-3}}\left(\frac{\Lambda_{N_f}}{a}\right)^{8-2 N_f}&{\rm
  for}&N_f \geq 1 ,
    \end{array}\right.
\end{eqnarray}
where $\Lambda_{N_f}^{8-2 N_f}$ is replaced with $q$ for $N_f=4$.
In the case when the hypermultiplets have non-vanishing bare masses,
the mass term is added to the action: $S_{mass}=-i \sum_{f=1}^{N_f}
\sum_{i=1}^2m_f
\tilde{\zeta}_{fi} \zeta_{fi}$. The instanton calculus is
performed straightforwardly. Using the equation,
\begin{eqnarray}
\sum_{i,j=1}^2\tilde{\zeta}^{i}\zeta_{i}
\tilde{\zeta}_{j}\zeta_{j}=0, \quad
\left(\sum_{i=1}^2\tilde{\zeta}^{i}\zeta_{i}\right)^2=
\left(\sum_{i=1}^2\tilde{\zeta}_{i}\zeta_{i}\right)^2,
\end{eqnarray}
we obtain the result;
\begin{eqnarray}
\label{massive}
  u_{2}=\frac{1}{2}a^2\left(\frac{\Lambda_{N_f}}{ a}\right)^{8-2N_f}
   \sum_{t_1=0}^{1} \cdots
  \sum_{t_{N_f}=0}^{1} \left(-\frac{1}{2}\right)^{N_f-M} I(N_f-M)
   a^{-2M}\prod_{g=1}^{N_f}m_g^{2 t_g},
\end{eqnarray}
where $M=\sum_{f=1}^{N_f}t_f$, and $\Lambda^{8-2 N_f}_{N_f}$ is
replaced by $q$ for $N_f=4$.
When $m_{N_f} \rightarrow \infty $ and $\Lambda_{N_f}\rightarrow 0$,
we obtain
\begin{eqnarray}
\label{nfsub1}
  u_{2}=\frac{1}{2}a^2
\frac{m_{N_f}^2\Lambda_{N_f}^{8-2N_f}}{ a^{8-2(N_f-1)}}
    \cdot \left(-\frac{1}{2}\right)^{N_f-1}I(N_f-1),
\end{eqnarray}
where we set the other masses to be zero.
{}From Eq.(\ref{vevu}) and (\ref{nfsub1}), we find that the result for
the massive theories
agrees with the decoupling relation: $m_{N_f}^2
\Lambda_{N_f}^{8-2N_f}=\Lambda_{N_f-1}^{8-2 (N_f-1)}$. For $N_f=4$,
the decoupling relation is given by $m^2_4 q = \Lambda_3^2$.

\reseteqnum
\section{Exact results versus instanton calculus}
In the previous section, we obtain the following results by the
instanton calculus for the
massless theories,
\begin{eqnarray}
\label{oneinst2}
&&  u_1=2 a^2 \times \left\{
    \begin{array}{ccc}
\vspace{1ex}
      \displaystyle
 \frac{1}{2^3} \left(\frac{\Lambda_0}{a}\right)^4&{\rm for}&N_f=0\,,\\
\vspace{1ex}
      \displaystyle
0&{\rm  for}&N_f \geq 1 ,
    \end{array}\right. \\
\label{instcal2}
&&  u_2=2 a^2 \times \left\{
    \begin{array}{lcc}
\vspace{1ex}
      \displaystyle
\frac{5}{2^9}\left(\frac{\Lambda_0}{a}\right)^8&{\rm for}&N_f=0\,,\\
\vspace{1ex}
      \displaystyle
(-1)^{N_f}\frac{2N_f-1}{2^{9-N_f} 3^{2
    N_f-3}}\left(\frac{\Lambda_{N_f}}{a}\right)^{8-2 N_f}&{\rm
  for}&N_f \geq 1 ,
    \end{array}\right.
\end{eqnarray}
and the decoupling relation is given by
\begin{eqnarray}
\label{decrel1}
  m_{N_f}^2
\Lambda_{N_f}^{8-2N_f}=\Lambda_{N_f-1}^{8-2 (N_f-1)}.
\end{eqnarray}
For $N_f=4$, $\Lambda_{N_f}^{8-2N_f}$ is replaced by $q$.
Note on the convention for $a$. The definition of $a$
differs between Eq.(\ref{oneinst}), Eq.(\ref{instcal}) and
Eq.(\ref{oneinst2}), Eq.(\ref{instcal2}) by factor 2.

On the other hand, from
Eq.(\ref{coeff}) and Eq.(\ref{u(a)}) the exact result predicts the
moduli $u(a)$ for the massless theories,
\begin{eqnarray}
\label{exact1}
&&  u_1=2 a^2 \times \left\{
    \begin{array}{ccc}
\vspace{1ex}
      \displaystyle
 \frac{1}{2^5} \left(\frac{\tilde{\Lambda}_0}{a}\right)^4&{\rm
   for}&N_f=0\,,\\
\vspace{1ex}
      \displaystyle
0&{\rm  for}&N_f \geq 1 ,
    \end{array}\right. \\
\label{exact2}
&&  u_2=2 a^2 \times \left\{
    \begin{array}{lcc}
\vspace{1ex}
      \displaystyle
\frac{5}{2^{13}}\left(\frac{\tilde{\Lambda}_0}{a}\right)^8&{\rm
  for}&N_f=0\,,\\
\vspace{1ex}
      \displaystyle
-\frac{3}{2^{12}}\left(\frac{\tilde{\Lambda}_1}{a}\right)^6&{\rm
  for}&N_f=1\,,\\
\vspace{1ex}
      \displaystyle
\frac{1}{2^{11}}\left(\frac{\tilde{\Lambda}_2}{a}\right)^4&{\rm
  for}&N_f=2\,,\\
\vspace{1ex}
      \displaystyle
\frac{1}{2^{10}}\left(\frac{\tilde{\Lambda}_3}{a}\right)^2&{\rm
  for}&N_f=3\,,\\
    \end{array}\right.
\end{eqnarray}
and the decoupling relation is given by
\begin{eqnarray}
  \label{decrel2}
&&m_{N_f}^2\tilde{\Lambda}_{N_f}^{8-2N_f}=\tilde{\Lambda}_{N_f-1}^{8-2
  (N_f-1)}\quad  {\rm for }\quad 1\leq N_f \leq 3 ,
\\
\label{decrel3}
&& \tilde{\Lambda}_3^2=64^2m_4^2 q \,.
\end{eqnarray}
To compare the exact result and the instanton calculus, we must relate
dynamical scales $\tilde{\Lambda}_{N_f}$ and $\Lambda_{N_f}$.
Only when $N_f$=0, the one-instanton contribution $u_1$ does not
vanish, and it is easily found from
Eq.(\ref{oneinst2}) and Eq.(\ref{exact1})  that the instanton
calculus is consistent with the exact result
in the one-instanton sector, if we identify the
dynamical scales
as $\tilde{\Lambda}_0=\sqrt{2}\Lambda_0$\cite{FP}.
In supersymmetric $SU(3)$ Yang-Mills theory, the instanton calculus
is consistent with
the instanton calculus in the one-instanton sector,
when the above relation  of the
dynamical scales of the $SU(2)$ theory holds\cite{IS}.

The two-instanton correction to $\VEV{u}$ is given by $u_2$.
According to the exact result, no quantum correction to $\VEV{u}$
exists for $N_f$ =4.
Using the relation $\tilde{\Lambda}_0=\sqrt{2}\Lambda_0$ and the
decoupling relations Eq.(\ref{decrel1}), (\ref{decrel2})
we obtain
the relation between the dynamical scales:
\begin{eqnarray}
  \label{dynamical}
  \tilde{\Lambda}^{8-2N_f}_{N_f}=16\Lambda^{8-2N_f}_{N_f}.
\end{eqnarray}
{}From Eq.(\ref{instcal2}), (\ref{exact2}) and (\ref{dynamical}), we find
that the microscopic instanton calculus
agrees
with the exact results
for $N_f=1,2$ as well as $N_f=0$.\footnote{Furthermore, for the
  massive $N_f=1,2$ theories, it can be
 shown that the instanton calculus Eq.(\ref{massive}) agrees with
 the exact results\cite{Ohta}.}
However we also find discrepancies between them for $N_f=3,4$.
For $N_f=3$, the difference of the moduli $u$ between the instanton
calculus and the
exact result is a constant.
For $N_f=4$, we find the quantum
correction to the moduli $u$, which disagrees with the assumption
used by \cite{SW}.

At first sight, the decoupling relations connecting the $N_f=3$ and
$N_f=4$ theories
give an  inconsistency between the exact result and the instanton
calculus. The relation of the instanton calculus $m_4^2
q=\Lambda_3^2$ and that of the exact result $64^2 m_4^2
q=\tilde{\Lambda}_3^2$ disagree with Eq.(\ref{dynamical}).
However, this inconsistency can be resolved by changing the
regularization scheme of the instanton calculus. We will discuss this
point in
more detail in appendix D.

In the similar way, it can be evaluated the four-point function
$\VEV{\bar{\lambda}\bar{\lambda}\bar{\psi}\bar{\psi}}$ by the
  instanton calculus\cite{FP,DKM,DKM2}, and we find that the
  non-trivial relation
  Eq.(\ref{u(a)}) holds for $N_f=1,2$ as well as $N_f=0$\cite{FT,DKM}.
For $N_f=3,4$, this four-point function does not depend on ${\cal F}_2$
and therefore it is not useful to check the exact result. For $N_f=4$,
the finite correction to the coupling constant is calculated in
\cite{DKM3}: $\tau_{eff}=\tau+(i/2\pi)\sum_{k}{\cal F}_k q^{k/2}$ and we
find that Eq.(\ref{u(a)}) with
$\Lambda_{N_f}^{8-2N_f}$ replaced by $q$ holds in this case:
\begin{eqnarray}
  u(a)=2 a^2\left\{1-\frac{1}{2}\sum_{k=1}^{\infty}k{\cal
    F}_k q^{k/2}\right\}=8\pi iq\frac{\partial {\cal F}(a)}{\partial q},
\end{eqnarray}
where the prepotential is given by ${\cal F}=\tau_{eff}a^2/2$ .

\section{Conclusion}
We have constructed the supersymmetric multi-instanton in $N=2$
supersymmetric QCD and derived the supersymmetric instanton action.
The instanton calculus has been performed in $N=2$ supersymmetric
QCD and we have found that the
instanton calculus agrees with the Seiberg-Witten's result for the
$N_f\leq2$ theories.
We have also found the discrepancies between them for $N_f=3,4$.
These results mean that the Seiberg-Witten solution should be
modified for the $N_f=3,4$ theories. For $N_f=3$ the curve is
determined except for a constant added to $u$,  since there is no discrete
symmetry in the $u$ plane. We can shift $u$ in the elliptic curve for
$N_f=3$ by the constant, so that the discrepancy for $N_f=3$ is
resolved.
We obtain the correct curve for the massive $N_f=3$ theory:
\begin{eqnarray}
&&y^2=x^2(x-u-\frac{1}{2^43^3}\tilde{\Lambda}_3^2)
-\frac{1}{64}\tilde{\Lambda}_3^2(x-u-\frac{1}{2^43^3}
\tilde{\Lambda}_3^2)^2-\frac{1}{64}(m_1^2+m_2^2+m_3^2)
\tilde{\Lambda}_3^2 (x-u-\frac{1}{2^43^3}\tilde{\Lambda}_3^2)
\nonumber
\\
&&\hspace{15ex}+\frac{1}{4}m_1m_2m_3\tilde{\Lambda}_3
x-\frac{1}{64}(m_1^2m_2^2+m_2^2m_3^2 +m_1^2m_3^2)\tilde{\Lambda}_3^2\,  .
\end{eqnarray}
This modification of the curve does not affect the $N_f<3$
theories. In the decoupling limit: $\tilde{\Lambda}_3\rightarrow
0$, $m_3\rightarrow \infty$ with $\tilde{\Lambda}_2^4=m_3^2
\tilde{\Lambda}_3^2$ fixed, we obtain the massive $N_f=2$ curve, which agrees
with the massive $N_f=2$ curve derived from the original $N_f=3$ curve.

We also need to solve the discrepancy for the massless $N_f=4$
theory. The quantum correction does exist in this case, contrary to
\cite{SW}. The prepotential is given by ${\cal F}(a)=\tau_{eff} a^2/2$
and the
moduli $u$ is given by
\begin{eqnarray}
  u=8\pi iq\frac{\partial {\cal F}(a)}{\partial q}=2a^2\frac{d
    \tau_{eff}}{d \tau},
\end{eqnarray}
\begin{eqnarray}
  a(u)=\frac{\sqrt{2u}}{2}\left(\frac{d \tau_{eff}}{d
      \tau}\right)^{-1/2},\quad a_D(u)=\tau_{eff} a(u).
\end{eqnarray}
The period of the meromorphic one-form on the Seiberg-Witten curve for
the massless
$N_f=4$ theory gives the above result by replacing $\tau$ and $u$ in
the elliptic curve with $\tau_{eff}$ and $(d\tau_{eff}/d
\tau)^{-1}u$, respectively:
\begin{eqnarray}
 y^2=\left(x^2-c_2\left(\tau_{eff}\right)^2
   u^2\left(\frac{d\tau_{eff}}{d\tau}\right)^{-2}\right)
\left(x-c_1\left(\tau_{eff}\right) u\left(\frac{d\tau_{eff}}
{d\tau}\right)^{-1}\right).
\end{eqnarray}

There remains an unsolved inconsistency between the instanton calculus
and the exact result for the massive $N_f=4$ theory.
The curve for the massive $N_f=4$ theory should be modified, such that
it satisfies  the desirable conditions in both $m_4\rightarrow 0$ and
$\infty$ limits.
The curve must go to the modified $N_f=3$ curve in the decoupling
limit: $m_4 \rightarrow \infty$ and $q\rightarrow 0$ with $m_4^2 q$
fixed and go to the modified massless $N_f=4$ curve in the
massless limit: $m_4 \rightarrow 0$.

We also have calculated the instanton corrections for $N_f>4$.
Although the theory is not asymptotically free, there exist the exact
results\cite{APSei} and the comparison between them may give some new
insights of the theory.

\vskip5cm
\centerline{\large\bf Acknowledgment}

\noindent
We thank K.~Ito, N.~Sasakura, N.~Dorey and
M.~P.~Mattis for discussions.
We would especially like to acknowledge the numerous
valuable discussions with H.~Aoyama, S.~Wada and S.~Sugimoto.
The work of T.H. is supported  in part by the Grant-in-Aid for
JSPS fellows.
\vskip 2cm
\centerline{\Large\bf Appendices}
\appendix
\section{Conventions}
In this section, we summarize our conventions used in this paper.
The sigma matrix is defined by
\begin{eqnarray}
(\sigma_{\mu})_{\alpha\dot{\beta}}=(\mbox{\boldmath$\sigma$},\,i),
\quad
(\bar{\sigma}_{\mu})^{\dot{\alpha}\beta}=(\mbox{\boldmath$\sigma$},\,-i),
\end{eqnarray}
\begin{eqnarray}
(\sigma_{\mu\nu})^{\dot{\alpha}}_{\,\,\dot{\beta}}=\frac{1}{4i}
(\bar{\sigma}_{\mu}\sigma_{\nu}-\bar{\sigma}_{\nu}\sigma_{\mu}),
\quad
(\bar{\sigma}_{\mu\nu})_{\alpha}^{\,\,\beta}=\frac{1}{4i}
(\sigma_{\mu}\bar{\sigma}_{\nu}-\sigma_{\nu}\bar{\sigma}_{\nu}),
\end{eqnarray}
where $\mbox{\boldmath$\sigma$}$ is the Pauli matrices.
The invariant tensor is defined by
\begin{eqnarray}
\epsilon^{12}=\epsilon_{21}=1,
\quad \epsilon_{12}=\epsilon^{21}=-1,
\end{eqnarray}
and we use the same contraction rule of the spinor index as \cite{WB}.
The quaternion $M$ given by
\begin{eqnarray}
&&M=M_1\hat{\imath}+M_2\hat{\jmath}+M_3\hat{k}+M_4,\nonumber\\
&&\hat{\imath}^2=\hat{\jmath}^2=\hat{k}^2=-1,\quad
\hat{\imath}\times\hat{\jmath}=\hat{k},
\end{eqnarray}
has the following matrix representation:
\begin{eqnarray}
M_{s\dot{s}}=-i(\sigma_{\mu})_{s\dot{s}}M_{\mu}.
\end{eqnarray}
We denote the quaternion conjugate of $M$ as $\bar{M}$.
Namely
\begin{eqnarray}
\bar{M}=-M_1\hat{\imath}-M_2\hat{\jmath}-M_3\hat{k}+M_4,
\end{eqnarray}
and it has the following matrix representation:
\begin{eqnarray}
\bar{M}^{\dot{r}r}=i(\bar{\sigma}_{\mu})^{\dot{r}r}M_{\mu}.
\end{eqnarray}
For a quaternion matrix $M_{IJ}$, we use the symbol $\dagger$ as the
transpose of the quaternion conjugate of $M_{IJ}$:
\begin{eqnarray}
(M^{\dagger})_{IJ}=\bar{M}_{JI}.
\end{eqnarray}
Since the instanton mixes the color and
spinor indices, it is convenient to use the dotted index for the
color index.
We denote the color index as
\begin{eqnarray}
{A_{\mu}}_{\,\,\dot{s}}^{\dot{r}},
\quad q^{\dot{r}},\quad \tilde{q}_{\dot{s}}.
\end{eqnarray}

\section{Lagrangian of $N=2$ supersymmetric $SU(2)$ QCD}
In Euclidean space we have the Lagrangian,
\begin{eqnarray}
&&{\cal L}=
{\rm tr}\left\{\frac{1}{2}F_{\mu\nu}F_{\mu\nu}
-2i\bar{\lambda}\Dbarslash\lambda-D^2\right.
+2(D_{\mu}\phi)^{\dagger}D_{\mu}\phi-2i\bar{\psi}\Dbarslash\psi-2F^{\dagger}F
\nonumber\\
&&\hspace{5ex}+2\sqrt{2}ig\left(\lambda[\psi,\phi^{\dagger}]
+\bar{\lambda}[\bar{\psi},\phi]\right)-2g
D[\phi^{\dagger},\phi]\biggl\}
\nonumber\\
&&\hspace{5ex}+(D_{\mu}Q)^{\dagger}D_{\mu}Q
+D_{\mu}\tilde{Q}(D_{\mu}\tilde{Q})^{\dagger}
 -i\bar{q}\Dbarslash q -i\bar{\tilde{q}}\Dbarslash\tilde{q}
 -F^{\dagger }_Q F_{Q}+F_{\tilde{Q}} F_{\tilde{Q}}^{\dagger}\\
\nonumber
&&\hspace{5ex}+\sqrt{2}ig\left(Q^{\dagger}\lambda
  q-\bar{q}\bar{\lambda}Q-\tilde{q}\lambda\tilde{Q}^{\dagger}
+\tilde{Q}\bar{\lambda}\bar{\tilde{q}}\right)+g\left(Q^{\dagger}DQ
-\tilde{Q}D\tilde{Q}^{\dagger}\right)\\
\nonumber
&&\hspace{5ex}+\sqrt{2}ig\left(\tilde{q}\phi q+\tilde{q}\psi Q+\tilde{Q}\psi
  q-\bar{q}\phi^{\dagger}\bar{\tilde{q}}-Q^{\dagger}\bar{\psi}\bar{\tilde{q}}
-\bar{q}\bar{\psi}\tilde{Q}^{\dagger}\right)\\
\nonumber
&&\hspace{5ex}-i\sqrt{2}g\left(F_{\tilde{Q}}\phi Q+\tilde{Q}FQ
+\tilde{Q}\phi F_{Q}-Q^{\dagger}\phi^{\dagger}F^{\dagger}_{\tilde{Q}}
-Q^{\dagger}F^{\dagger}\tilde{Q}^{\dagger}
-F^{\dagger}_Q\phi^{\dagger}\tilde{Q}^{\dagger}\right),
\end{eqnarray}
where color, flavor and  spin indices are suppressed.
The covariant derivatives and the field strength are defined by,
\begin{eqnarray}
  \label{covderiv}
&&D_{\mu}\phi=\partial_{\mu} \phi-ig[A_{\mu},\phi],\\
&&D_{\mu}Q=\left(\partial_{\mu}-igA_{\mu}\right)Q,\quad D_{\mu}\tilde{Q}
=\left(\partial_{\mu}\tilde{Q}+ig\tilde{Q}A_{\mu}\right),\\
&& F_{\mu\nu}=\partial_{\mu}A_{\nu}-\partial_{\nu}A_{\mu}-ig[A_{\mu},A_{\nu}].
\end{eqnarray}
We define the $N=2$ supersymmetric mass term by,
\begin{eqnarray}
  \label{massterm}
  {\cal L}_m=-i\sum_{f=1}^{N_f}
  m_f\left(-\tilde{q}_fq_f+F_{\tilde{Q}f}Q_f+\tilde{Q}_f
    F_{Qf}\right)+h.c.\quad .
\end{eqnarray}
In this convention, $(\lambda,\psi)$ and  $(Q,\tilde{Q}^{\dagger})$ are
$SU(2)_R $ doublets. We adopt the superpotential:
$-\sqrt{2}i\tilde{Q}\Phi Q-i\sum_{f=1}^{N_f}m_f\tilde{Q}_fQ_f$, which
coincides with one used in \cite{SW} by replacing
$-i\tilde{Q}_f\rightarrow \tilde{Q}_f$.

\section{Supersymmetry}
The action has $N=2$ supersymmetry. One of the super transformations is
given by,
\begin{eqnarray}
&&\delta\phi=-\sqrt{2}i\xi\psi\nonumber\\
&&\delta\phi^{\dagger}=\sqrt{2}i\bar{\xi}\bar{\psi}\nonumber\\
&&\delta\psi=-\sqrt{2}i\xi F-\sqrt{2}\sigma_{\mu}\bar{\xi}D_{\mu}\phi
\nonumber\\
&&\delta\bar{\psi}=-\sqrt{2}\xi\sigma_{\mu}D_{\mu}\phi^{\dagger}
+\sqrt{2}i\bar{\xi}F^{\dagger}\nonumber\\
&&\delta F
=\sqrt{2}\bar{\xi}_{\dot{\alpha}}\left\{(\Dbarslash\psi)^{\dot{\alpha}}
+\sqrt{2}g[\bar{\lambda}^{\dot{\alpha}},\phi]\right\}
\\
&&\delta F^{\dagger
}=\sqrt{2}\xi^{\alpha}\left\{(\Dslash\bar{\psi})_{\alpha}
+\sqrt{2}g[\phi^{\dagger },\lambda_{\alpha}]\right\}
\nonumber\\
&&\delta A_{\mu}
=\bar{\lambda}\bar{\sigma}_{\mu}\xi+\bar{\xi}\bar{\sigma}_{\mu}\lambda
\nonumber\\
&&\delta\lambda=-\bar{\sigma}_{\mu\nu}\xi F_{\mu\nu}+\xi D
\nonumber\\
&&\delta\bar{\lambda}=-\bar{\xi}\sigma_{\mu\nu}F_{\mu\nu}+\bar{\xi}D
\nonumber\\
&&\delta D=i\xi\Dslash\bar{\lambda}-i\bar{\xi}\Dbarslash\lambda
\nonumber
\end{eqnarray}
for the vector multiplets and
\begin{eqnarray}
&&\delta Q=-\sqrt{2}i\xi q
\nonumber\\
&&\delta Q^{\dagger}=\sqrt{2}i\bar{\xi}\bar{q}
\nonumber\\
&&\delta q=-\sqrt{2}i\xi F_{Q}-\sqrt{2}\sigma_{\mu}\bar{\xi}D_{\mu}Q
\nonumber\\
&&\delta\bar{q}
=-\sqrt{2}\xi\sigma_{\mu}D_{\mu}Q^{\dagger}+\sqrt{2}i\bar{\xi}
F_{Q}^{\dagger}
\nonumber\\
&&\delta F_{Q}
=\sqrt{2}\bar{\xi}\bar{\sigma}_{\mu}D_{\mu}q
+2g\bar{\xi}\bar{\lambda}Q
\nonumber\\
&&\delta F^{\dagger }_{Q}
=\sqrt{2}D_{\mu}\bar{q}\bar{\sigma}_{\mu}\xi+
2gQ^{\dagger }\xi\lambda
\\
&&\delta\tilde{Q}=-\sqrt{2}i\xi\tilde{q}
\nonumber\\
&&\delta\tilde{Q}^{\dagger}=\sqrt{2}i\bar{\xi}\bar{\tilde{q}}
\nonumber\\
&&\delta\tilde{q}=-\sqrt{2}i\xi F_{\tilde{Q}}
-\sqrt{2}\sigma_{\mu}\bar{\xi} D_{\mu}\tilde{Q}
\nonumber\\
&&\delta\bar{\tilde{q}}=-\sqrt{2}\xi\sigma_{\mu} D_{\mu}\tilde{Q}^{\dagger}
+\sqrt{2}i\bar{\xi}F^{\dagger}_{\tilde{Q}}
\nonumber\\
&&\delta F_{\tilde{Q}}
=\sqrt{2}\bar{\xi}\bar{\sigma}_{\mu}D_{\mu}\tilde{q}
-2g\tilde{Q}\bar{\xi}\bar{\lambda}
\nonumber\\
&&\delta F_{\tilde{Q}}^{\dagger }
=\sqrt{2} D_{\mu}\bar{\tilde{q}}\sigma_{\mu}\xi
-2g\xi\lambda \tilde{Q}^{\dagger }
\nonumber
\end{eqnarray}
for the hypermultiplets. The other super transformation is given by
$SU(2)_R $ rotation of this transformation.

\section{Regularization scheme}
In this section, we discuss the regularization scheme. In section 5, we
define the instanton measure so that the decoupling relation agrees
with the $\overline{DR}$ scheme\cite{FP}. In the $\overline{DR}$ scheme, the
dynamical scales of the low and high energy theories relate by the following
decoupling relation:
\begin{eqnarray}
\label{drbar}
  \left(\frac{\Lambda_{0L}}{m}\right)^{b_{0L}}
=\left(\frac{\Lambda_{0H}}{m}\right)^{b_{0H}}
{\rm for\,\, any\,\, representation},
\end{eqnarray}
where $b_0$ is the coefficient of the $\beta$ function and $m$ is the
mass of the matter decoupled in the low energy theory. For $N=2$
supersymmetric $SU(2)$ QCD, $b_0=4-N_f$.
We define the dynamical scale by $\Lambda^{b_0}=\mu^{b_0}e^{-8\pi^2/g^2(\mu)}$.

We can use the another scheme in the instanton calculus. For example,
we will adopt a scheme, in which the decoupling relation is given by
\begin{eqnarray}
&&  \left(\frac{\alpha\Lambda_{0L}}{
      m}\right)^{b_{0L}}=\left(\frac{\alpha\Lambda_{0H}}{m}\right)^{b_{0H}}
{\rm for\,\, \,\, adjoint\,\, representation},\\
&&
\left(\frac{\Lambda_{0L}}{m}\right)^{b_{0L}}
=\left(\frac{\Lambda_{0H}}{m}\right)^{b_{0H}}
{\rm for\,\, fundamental \,\, representation},
\end{eqnarray}
where $\alpha$ is a numerical constant.
In this scheme, the $k$-instanton measure of the $SU(N_c)$ vector
multiplet is given by multiplying
$\alpha^{2N_c k}$
to that of the $\overline{DR}$ scheme. In this scheme, the result of the
instanton
calculus is the following:
\begin{eqnarray}
&&  u_1=2 a^2 \times \left\{
    \begin{array}{ccc}
\vspace{1ex}
      \displaystyle
 \frac{\alpha^4}{2^3} \left(\frac{\Lambda_0}{a}\right)^4&{\rm for}&N_f=0\,,\\
\vspace{1ex}
      \displaystyle
0&{\rm  for}&N_f \geq 1 ,
    \end{array}\right. \\
&&  u_2=2 a^2 \times \left\{
    \begin{array}{lcc}
\vspace{1ex}
      \displaystyle
\frac{5\alpha^8}{2^9}\left(\frac{\Lambda_0}{a}\right)^8&{\rm for}&N_f=0\,,\\
\vspace{1ex}
      \displaystyle
(-1)^{N_f}\frac{(2N_f-1)\alpha^8}{2^{9-N_f} 3^{2
    N_f-3}}\left(\frac{\Lambda_{N_f}}{a}\right)^{8-2 N_f}&{\rm
  for}&N_f \geq 1 .
    \end{array}\right.
\end{eqnarray}

The one-instanton calculus is consistent  with the
exact results Eq.(\ref{exact1}), when we identify the
dynamical scales as $\tilde{\Lambda}_0=\sqrt{2}\alpha \Lambda_0$.
By the decoupling relation for the fundamental representation, we
obtain the relation between the dynamical scales,
\begin{eqnarray}
\label{appendyn}
  \tilde{\Lambda}_{N_f}^{8-2N_f}=16 \alpha^8 \Lambda_{N_f}^{8-2N_f}.
\end{eqnarray}
For any $\alpha$ the two-instanton calculus and the exact results agree for
$N_f=0,1,2$ and there appear the discrepancies for $N_f=3,4$, as
we have already seen in section 6.
{}For the supersymmetric $SU(3)$ Yang-Mills theory, the instanton
calculus also agrees with the exact results in the one-instanton
sector for any $\alpha$.
Therefore our result does not depend on the regularization scheme.
The decoupling relations connecting the $N_f=3$ and $N_f=4$
theories are given by $\Lambda_3^2=m_4^2 q$ for the instanton calculus
and $\tilde{\Lambda}_3^2=64^2 m_4^2 q$ for the exact results. These
relations are consistent with Eq.(\ref{appendyn}), if we choose
$\alpha=2$.

\newcommand{\J}[4]{{\sl #1} {\bf #2} (19#3) #4}
\newcommand{\MPL}{Mod.~Phys.~Lett.}
\newcommand{\NP}{Nucl.~Phys.}
\newcommand{\PL}{Phys.~Lett.}
\newcommand{\PR}{Phys.~Rev.}
\newcommand{\PRL}{Phys.~Rev.~Lett.}
\newcommand{\AP}{Ann.~Phys.}
\newcommand{\CMP}{Commun.~Math.~Phys.}
\newcommand{\CQG}{Class.~Quant.~Grav.}
\newcommand{\PRP}{Phys.~Rept.}
\newcommand{\SPU}{Sov.~Phys.~Usp.}
\newcommand{\RMPA}{Rev.~Math.~Pur.~et~Appl.}
\newcommand{\SPJ}{Sov.~Phys.~JETP}
\newcommand{\MP}{Int.~Mod.~Phys.}
\newcommand{\ZP}{Z.~Phys.}


\begin{thebibliography}{99}
\bibitem{SW} N.~Seiberg and E.~Witten, \J{\NP}{B426}{94}{19};\\
N.~Seiberg and E.~Witten, \J{\NP} {B431}{94}{484}.
\bibitem{KLYT} A.~Klemm, W.~Lerche, S.~Yankielowicz and S.~Theisen,
\J{\PL}{B344}{95}{169}.
\bibitem{AF} P.~C.~Argyres and A.~E.~Faraggi, \J{\PRL}{74}{95}{3931}.
\bibitem{HO}A.~Hanany and Y.~Oz, \J{\NP}{B452}{95}{283}.
\bibitem{APS}P.~C.~Argyres, M.~R.~Plesser and A.~D.~Shapere,
  \J{\PRL}{75}{95}{1699}.
\bibitem{AS}P.~C.~Argyres and A.~D.~Shapere, \J{\NP}{B461}{96}{437}.
\bibitem{AD}P.~C.~Argyres and M.~R.~Douglas, \J{\NP}{B448}{95}{93}.
\bibitem{APSW}P.~C.~Argyres, M.~R.~Plesser, N.~Seiberg and E.~Witten,
  \J{\NP}{B461}{71}{96}.
\bibitem{EHIY}T.~Eguchi, K.~Hori, K.~Ito and S.-K.~Yang, hep-th/9603002.
\bibitem{ADS}I.~Affleck, M.~Dine and N.~Seiberg, \J{\NP}{B241}{84}{493}.
\bibitem{NSVZ}V.~A.~Novikov, M.~A.~Shifman, A.~I.~Vainshtein and
  V.~I.~Zakharov, \J{\NP}{B260}{85}{157}.
\bibitem{AKMRV}D.~Amati, K.~Konishi, Y.~Meurice, G.~C.~Rossi and
G.~Veneziano, \J{\PRP}{162}{88}{169}.
\bibitem{FP}D.~Finnell and P.~Pouliot, \J{\NP}{B453}{95}{225}.
\bibitem{DKM}N.~Dorey, V.~V.~Khoze and M.~P.~Mattis, hep-th/9603136;\\
N.~Dorey, V.~V.~Khoze and M.~P.~Mattis, hep-th/9606199.
\bibitem{Yung} A.~Yung, hep-th/9605096.
\bibitem{FT} F.~Fucito and G.~Travaglini, hep-th/9605215.
\bibitem{IS} K.~Ito and N.~Sasakura, \J{\PL}{B382}{96}{95}.
\bibitem{DKM2}N.~Dorey, V.~V.~Khoze and M.~P.~Mattis, hep-th/9607066.
\bibitem{AHSW}H.~Aoyama, T.~Harano, M.~Sato and S.~Wada,
  hep-th/9607076.
\bibitem{DKM3}N.~Dorey, V.~V.~Khoze and M.~P.~Mattis, hep-th/9607202.
\bibitem{IY}K.~Ito and S.-K.~Yang, \J{\PL}{B366}{96}{165}.
\bibitem{Matone}M.~Matone, \J{\PL} {B357}{95}{342}.
\bibitem{STY}J.~Sonnenschein, S.~Theisen and S.~Yankielowicz,
  hep-th/9510129.
\bibitem{EY}T.~Eguchi and S.-K.~Yang, hep-th/9510183.
\bibitem{Affleck}I.~Affleck, \J{\NP}{B191}{81}{429}.
\bibitem{AHSW2}H.~Aoyama, T.~Harano, M.~Sato and S.~Wada,
\J{\MPL}{A11}{96}{43}; \\
H.~Aoyama, T.~Harano, M.~Sato and S.~Wada,
\J{\NP}{B466}{96}{127}.
\bibitem{tHooft}G.~'t~Hooft, \J{\PR}{D14}{76}{3432}.
\bibitem{ADHM} M.~F.~Atiyah, V.~G.~Drinfeld, N.~J.~Hitchin and
Yu.~I.~Manin, \J{\PL}{A65}{78}{185};
V.~G.~Drinfeld and Yu.~I.~Manin, \J{\CMP}{63}{78}{177}.
\bibitem{CWS}N.~H.~Christ, E.~J.~Weinberg and
  N.~K.~Stanton, \J{\PR}{D18}{78}{2013}.
\bibitem{CGT} E.~Corrigan, P.~Goddard and S.~Templeton,
 \J{\NP}{B151}{79}{93}.
\bibitem{OCFGT}H.~Osborn, \J{\NP}{B140}{78}{45}; \\
E.~Corrigan, D.~Fairlie, P.~Goddard and S.~Templeton, \J{\NP}{B140}{78}{31}.
\bibitem{Osborn}H.~Osborn, \J{\AP}{135}{81}{373}.
\bibitem{Ohta}Y.~Ohta, hep-th/9604051, hep-th/9604059.
\bibitem{WB}J.~Wess and J.~Bagger, ~Supersymmetry and
  supergravity. Princeton; Princeton University Press 1983.
\bibitem{APSei}P.~C.~Argyres, M.~R.~Plesser and N.~Seiberg,
  hep-th/9603042
\end{thebibliography}
\end{document}